\documentclass[12pt]{imsart}

\usepackage[OT1]{fontenc}
\usepackage{url}
\usepackage{natbib,graphicx,amsmath,amssymb,fullpage,dsfont}
\usepackage{amsfonts}
\usepackage{fancybox}

\usepackage{lscape}
\usepackage{caption, subcaption}

\usepackage[linesnumbered,ruled]{algorithm2e}

\usepackage{comment}
\usepackage{color}
\usepackage{imsart}
\usepackage[colorlinks,citecolor=blue,urlcolor=blue]{hyperref}

\usepackage{tikz,enumerate}
\usepackage{multirow,ctable}
\usepackage[normalem]{ulem}
\usepackage[framemethod=TikZ]{mdframed}
\usetikzlibrary{arrows}

\newdimen\arrowsize
\pgfarrowsdeclare{arcsq}{arcsq}
{
  \arrowsize=0.2pt
  \advance\arrowsize by .5\pgflinewidth
  \pgfarrowsleftextend{-4\arrowsize-.5\pgflinewidth}
  \pgfarrowsrightextend{.5\pgflinewidth}
}
{
  \arrowsize=1.5pt
  \advance\arrowsize by .5\pgflinewidth
  \pgfsetdash{}{0pt} 
  \pgfsetroundjoin   
  \pgfsetroundcap    
  \pgfpathmoveto{\pgfpoint{0\arrowsize}{0\arrowsize}}
  \pgfpatharc{-90}{-140}{4\arrowsize}
  \pgfusepathqstroke
  \pgfpathmoveto{\pgfpointorigin}
  \pgfpatharc{90}{140}{4\arrowsize}
  \pgfusepathqstroke
}

\usepackage{paralist}
\newtheorem{proposition}{Proposition}

\newenvironment{proof}[1][. ]{{\bf Proof#1}}{\hfill$\square$\vskip\baselineskip}

\def\ES{{\mathcal E}}        

\newcommand{\R}{\mathbb{R}}

\newcommand{\C}{{\mathcal{C}}}
\newcommand{\D}{{\mathcal{D}}}

\newcommand{\card}{{\mbox{card}}}
\def\minimize#1#2{  {\underset{#1}{\mathrm{minimize}}}\left\{#2\right\}}
\def\argmin#1#2{  {\underset{#1}{\mathrm{argmin}}}\left\{#2\right\}}
\def\minn#1#2{  {\underset{#1}{\mathrm{min}}}\left\{#2\right\}}

\newcommand\numberthis{\addtocounter{equation}{1}\tag{\theequation}}

\newcounter{xxx}
\begin{document}

\begin{frontmatter}
\title{Exact Spike Train Inference\\  
 Via $\ell_0$ Optimization}
\runtitle{Spike Inference Via $\ell_0$ Optimization}

\vspace{5mm}

\begin{aug}
\author{Sean Jewell\ead[label = e1]{swjewell@uw.edu}} 
\address{Department of Statistics \\ University of Washington \\ Seattle, Washington 98195 \\ \printead{e1}}

\author{Daniela Witten\ead[label = e2]{dwitten@uw.edu}}
\address{Departments of Biostatistics and Statistics \\ University of Washington \\ Seattle, Washington 98195 \\ \printead{e2}}

\runauthor{S. Jewell and D. Witten}

\end{aug}

\begin{abstract}
In recent years, new  technologies in neuroscience have made it possible to measure the activities of large numbers of neurons simultaneously in behaving animals. For each neuron, a \emph{fluorescence trace} is measured; this can be seen as a first-order approximation of the neuron's activity over time. Determining the exact time at which a neuron spikes on the basis of its fluorescence trace is an important open problem in the field of computational neuroscience. 

Recently, 
a convex optimization problem  involving an $\ell_1$ penalty was proposed for this task. 
In this paper, we slightly modify that recent proposal by replacing the $\ell_1$ penalty with an $\ell_0$ penalty. In stark contrast to the conventional wisdom that $\ell_0$ optimization problems are computationally intractable, we show that the resulting optimization problem can be efficiently  solved for the global optimum using an extremely simple and efficient dynamic programming algorithm. Our \verb=R=-language implementation of the proposed algorithm runs in a few minutes on fluorescence traces of $100,000$ timesteps. Furthermore, our proposal leads to substantial improvements over the previous $\ell_1$ proposal, in simulations as well as  on two  calcium imaging data sets. 

\verb=R=-language software for our proposal is available on \verb=CRAN= in the package \\ \verb=LZeroSpikeInference=. Instructions for running this software in \verb=python= can be found at \url{https://github.com/jewellsean/LZeroSpikeInference}.

\end{abstract}


%

\end{frontmatter}

\maketitle
 

\section{Introduction}
\label{sec:intro}

  When  a neuron spikes, calcium floods the cell. 
   In order to quantify intracellular calcium levels, calcium imaging techniques make use of 
    fluorescent calcium indicator molecules \citep{dombeck2007imaging,ahrens2013whole,prevedel2014simultaneous}. 
    Thus, a  
  neuron's 
  \emph{fluorescence trace} can be seen as a first-order approximation of its activity level over time.

    However, the fluorescence trace itself is typically not of primary scientific interest: 
     instead, it is of interest to determine the underlying neural activity, that is, the specific times at which the neuron spiked. 
  Inferring the spike times on the basis of a fluorescence trace
   amounts to a challenging  deconvolution problem, which has been the focus of substantial investigation
 \citep{grewe2010high,pnevmatikakis2013bayesian,theis2016benchmarking,deneux2016accurate,sasaki2008fast,vogelstein2009spike,yaksi2006reconstruction,vogelstein2010fast,holekamp2008fast, friedrich2016fast, friedrichfast2017}. 
 In this paper, we propose a new approach for this task, which is based upon the following insight:  an auto-regressive model for calcium dynamics that has been extensively studied in the neuroscience literature \citep{vogelstein2010fast,friedrich2016fast,friedrichfast2017} leads directly to a simple $\ell_0$ optimization problem, for which an efficient and exact algorithm is available.

  
  \subsection{An Auto-Regressive Model for Calcium Dynamics}


In this paper, we will revisit an auto-regressive model for calcium dynamics that has been considered by a number of authors in the recent literature \citep{vogelstein2010fast,pnevmatikakis2016simultaneous,friedrich2016fast,friedrichfast2017}. 
%
We closely follow the notation of \cite{friedrichfast2017}. 
This model posits that
  $y_t$, the fluoresence at the $t$th timestep, is a noisy realization of $c_t$, the unobserved underlying calcium concentration at the $t$th timestep.
   In the absence of a spike at the $t$th timestep ($s_t=0$), the calcium concentration decays according to a $p$th-order auto-regressive process. However, if a spike occurs at the $t$th timestep ($s_t>0$), then the calcium concentration increases. Thus, 
\begin{align}
& y_t = \beta_0  + \beta_1 c_t + \epsilon_t, \quad \epsilon_t \sim_\text{ind.} (0, \sigma^2),  \quad t = 1, \ldots, T; \nonumber\\
& c_t = \sum_{i=1}^{p} \gamma_i c_{t-i} + s_t, \quad t = p + 1, \ldots, T.
\label{eq:model}
\end{align}
 In \eqref{eq:model},  
the quantities $\gamma_1,\ldots,\gamma_p$ are the parameters in the auto-regressive model. 
 Note that the quantity $y_t$  in \eqref{eq:model} is observed; all other quantities  are unobserved. 
   Since we would like to know whether a spike occurred at the $t$th timestep, the parameter of interest is $s_t$. 
     Figure~\ref{fig:overshrink}(a) displays a small dataset generated according to \eqref{eq:model}.
     
     In what follows, for ease of exposition,  we assume $\beta_0=0$ and $\beta_1=1$ in \eqref{eq:model}. This assumption is made without loss of generality, since $\beta_0$ and $\beta_1$ can be estimated from the data, and the observed fluorescence $y_1,\ldots,y_T$ centered and scaled accordingly. See Section~\ref{sec:extensions} for additional details.

        \cite{vogelstein2010fast}, \cite{friedrich2016fast},   and \cite{friedrichfast2017} seek to interpret $s_t$ in \eqref{eq:model} as the \emph{number} of spikes at the $t$th timestep. Thus, in principle it would be desirable to use   a  count-valued  distribution, such as the Poisson distribution, as a prior on $s_t$. However, because maximum a posteriori estimation of $s_t$ in \eqref{eq:model} using a Poisson distribution is computationally intractable, they instead suppose that $s_t$ has  an exponential distribution  \citep{vogelstein2010fast}. In the case of the first-order auto-regressive process ($p=1$), this leads \cite{vogelstein2010fast} to the optimization problem 
     \begin{equation}
\minimize{c_1,\ldots,c_T, s_2,\ldots,s_T}{  \frac{1}{2} \sum_{t=1}^T \left( y_t -  c_t \right)^2 + \lambda \sum_{t=2}^T | s_t |} \mbox{ subject to } s_t = c_t - \gamma c_{t-1} \geq 0, 
   \label{eq:convex}
   \end{equation}
   where $\lambda$ is a nonnegative tuning parameter that controls the trade-off between the fit of the estimated calcium to the observed fluorescence, and the sparsity of the estimated spike vector $\hat{s}_2,\ldots,\hat{s}_T$.  
    \cite{friedrich2016fast} and \cite{friedrichfast2017} instead consider a closely-related problem that results from including an additional $\ell_{1}$ penalty for the initial calcium concentration,
     \begin{equation}
\minimize{c_1,\ldots,c_T, s_2,\ldots,s_T}{  \frac{1}{2} \sum_{t=1}^T \left( y_t - c_t \right)^2 + \lambda |c_{1}| + \lambda \sum_{t=2}^T | s_t |} \mbox{ subject to } s_t = c_t - \gamma c_{t-1} \geq 0. 
   \label{eq:convex-fried}
   \end{equation}
Both \eqref{eq:convex} and \eqref{eq:convex-fried} are convex optimization problems, which can be solved for the global optimum using a well-developed set of optimization algorithms 
 \citep{boydconvex,elemstatlearn,hastie2015statistical,bien2016penalized}.  
  Because $\hat{s}_2,\ldots,\hat{s}_T$ are not integer-valued, they cannot be directly interpreted as the number of spikes at each timestep; however, informally, a larger value of $\hat{s}_t$ can be interpreted as indicating greater certainty that one or more spikes  occurred at the $t$th timestep.

     In this paper, we re-consider the model \eqref{eq:model} that originally motivated the optimization problems \eqref{eq:convex} and \eqref{eq:convex-fried} in the recent literature \citep{vogelstein2010fast,friedrich2016fast,friedrichfast2017}. Rather than interpreting $s_t$ in \eqref{eq:model} as the number of spikes at the $t$th timestep, we interpret its sign as an  indicator for whether or not at least one spike occurred: that is, $s_t=0$ indicates no spikes at the $t$th timestep, and $s_t>0$ indicates the occurrence of at least one spike. Then, in the case of a first-order auto-regressive model ($p=1$), \eqref{eq:model} leads naturally to the optimization problem 
   \begin{equation}
\minimize{c_1,\ldots,c_T, s_2,\ldots,s_T}{  \frac{1}{2} \sum_{t=1}^T \left( y_t -  c_t \right)^2 + \lambda \sum_{t=2}^T 1_{\left( s_t \neq 0 \right) }} \mbox{ subject to } s_t = c_t - \gamma c_{t-1} \geq 0,
   \label{eq:nonconvex-pis1}
   \end{equation}
   where $1_{(A)}$  is an indicator variable that equals $1$ if the event $A$ holds, and $0$ otherwise.  In \eqref{eq:nonconvex-pis1}, $\lambda$ is a non-negative tuning 
   parameter that controls the trade-off between the fit of the estimated calcium to the observed fluorescence, and the number of timesteps at which a spike is estimated to  occur. 
 
  Unfortunately, the optimization problem \eqref{eq:nonconvex-pis1} 
    is highly non-convex, due to the presence of the indicator variable.  
    In the statistics literature, this term is known as an $\ell_0$ penalty. It is well-known that optimization 
    involving $\ell_0$ penalties is typically computationally intractable: in general, no efficient algorithms are available to solve for the global optimum. 
    
In fact,  the convex optimization problem \eqref{eq:convex} considered in  \cite{vogelstein2010fast}, and its close cousin \eqref{eq:convex-fried} considered in \cite{friedrich2016fast} and \cite{friedrichfast2017}, can be viewed as convex relaxations to the problem \eqref{eq:nonconvex-pis1}. That is, if we replace the $\ell_0$ penalty in \eqref{eq:nonconvex-pis1} with an $\ell_1$ penalty, then that we arrive exactly at the problem \eqref{eq:convex}.

\subsection{Contribution of This Paper}
\label{subsec:contributions}

In the previous subsection, we established that the
 optimization problems \eqref{eq:convex} and \eqref{eq:convex-fried}, studied by \cite{vogelstein2010fast}, \cite{friedrich2016fast}, and \cite{friedrichfast2017}, can be seen as convex relaxations of the $\ell_0$ optimization problem \eqref{eq:nonconvex-pis1}, which follows directly from the model \eqref{eq:model}. 
 In fact, under the model \eqref{eq:model}, \eqref{eq:nonconvex-pis1} is the ``right" optimization problem to be solving; 
 \eqref{eq:convex} and \eqref{eq:convex-fried}  are simply computationally tractable approximations to this problem. 
  (In fact,  \cite{friedrichfast2017} allude to this in the ``Hard shrinkage and $\ell_{0}$ penalty'' section of their paper.)  

\begin{figure}[h]
\begin{center}
(a) \hspace{27mm} (b) \hspace{27mm} (c) \hspace{27mm} (d) \hspace{27mm} (e) \hspace{12mm}
\includegraphics[scale = 0.55,clip]{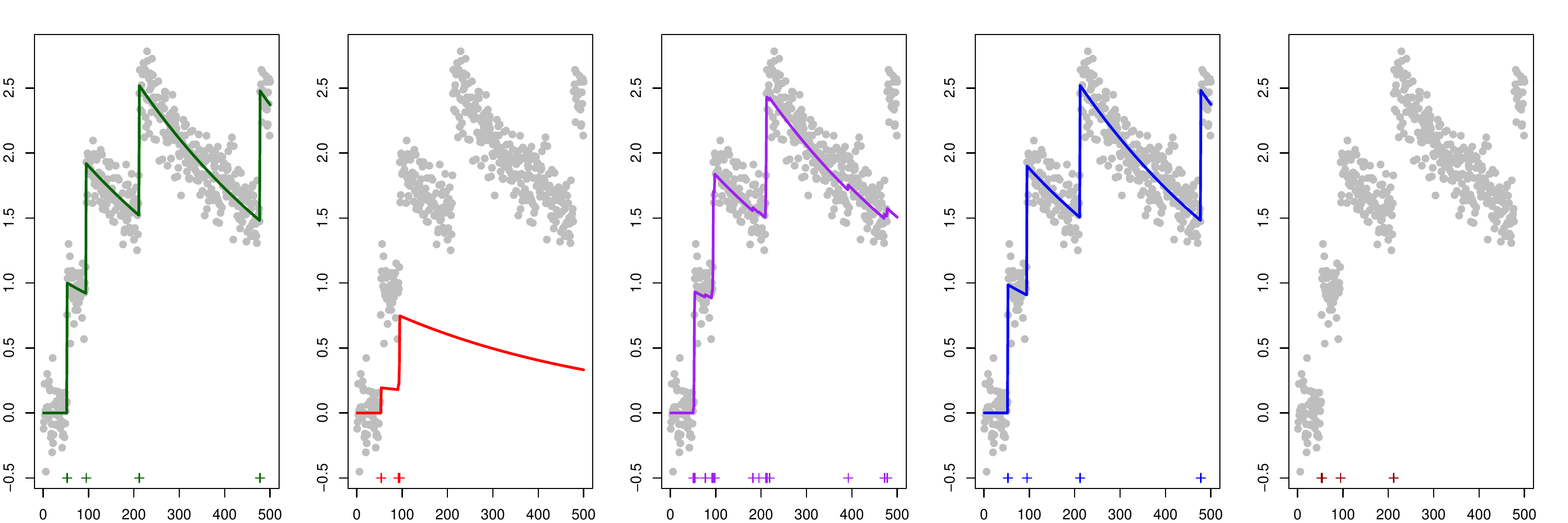}
\caption{A toy simulated data example. In each panel, the $x$-axis represents time. Observed fluorescence values are displayed in (\protect\includegraphics[height=0.4em]{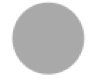}). 
\emph{(a)}: Unobserved calcium concentrations  (\protect\includegraphics[height=0.4em]{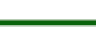}) and true spike times  (\protect\includegraphics[height=0.4em]{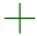}). Data were generated according to the model  \eqref{eq:model}. \emph{(b)}: Estimated calcium concentrations (\protect\includegraphics[height=0.4em]{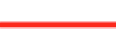}) and spike times (\protect\includegraphics[height=0.4em]{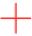}) that result from solving the $\ell_1$ optimization problem \eqref{eq:convex-fried} with the value of $\lambda$ that yields the true number of spikes. This value of $\lambda$ leads to very poor estimation of both the underlying calcium dynamics and the spikes. \emph{(c)}: Estimated calcium concentrations (\protect\includegraphics[height=0.4em]{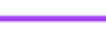}) and spike times (\protect\includegraphics[height=0.4em]{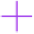}) that result from solving the $\ell_1$ optimization problem \eqref{eq:convex-fried} with the largest value of $\lambda$ that results in at least one estimated spike within the vicinity of each true spike. This value of $\lambda$ results in 19 estimated spikes, which is far more than the true number of spikes. The poor performance of the $\ell_1$ optimization problem in panels (b) and (c) is a consequence of the fact that the $\ell_1$ penalty performs shrinkage as well as spike estimation; this is discussed further in Section~\ref{subsec:contributions}. \emph{(d)}: Estimated calcium concentrations (\protect\includegraphics[height=0.4em]{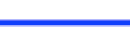}) and spike times (\protect\includegraphics[height=0.4em]{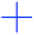}) that result from solving the $\ell_0$ optimization problem \eqref{eq:nonconvex-nopos}. \emph{(e)} The four spikes in panel (c) associated with the largest estimated increase in calcium (\protect\includegraphics[height=0.4em]{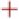}); we refer to this in the text as the post-thresholding $\ell_1$ estimator. Since the estimated calcium is not well-defined after post-thresholding, we do not plot the estimated calcium concentration.}
\label{fig:overshrink}
\end{center}
\end{figure}

However, using an $\ell_1$ norm to approximate an $\ell_0$ norm comes with computational advantages at the expense of substantial performance disadvantages: in particular, the use of an $\ell_1$ penalty tends to \emph{overshrink} the fitted estimates \citep{adaptivelasso06}.  This can be seen quite clearly in Figures~\ref{fig:overshrink}(b) and \ref{fig:overshrink}(c). 
Retaining only the four spikes in Figure~\ref{fig:overshrink}(c) associated with the largest increases in calcium leads to an improvement in spike detection (Figure~\ref{fig:overshrink}(e); this is referred to as the \emph{post-thresholding $\ell_1$ estimator} in what follows), but still one of the four true spikes is missed. 



%
  
In this paper, we consider a slight modification of \eqref{eq:nonconvex-pis1} that results from removing the positivity constraint,
  \begin{equation}
\minimize{c_1,\ldots,c_T,s_2,\ldots,s_T}{  \frac{1}{2} \sum_{t=1}^T \left( y_t - c_t \right)^2 + \lambda \sum_{t=2}^T 1_{\left( s_{t} \neq 0\right) }} \mbox{ subject to } s_t = c_t - \gamma c_{t-1}. 
   \label{eq:nonconvex-nopos}
   \end{equation}
In practice, the distinction between the problems \eqref{eq:nonconvex-nopos} and \eqref{eq:nonconvex-pis1} is quite minor: on real data applications, for appropriate choices of the decay rate $\gamma$, the solution to \eqref{eq:nonconvex-nopos} tends to satisfy the constraint in \eqref{eq:nonconvex-pis1}, and so the solutions are identical.

Like problem \eqref{eq:nonconvex-pis1}, solving problem \eqref{eq:nonconvex-nopos} for the global optimum appears, at a glance, to be computationally intractable --- we (the authors) are only aware of a few $\ell_0$ optimization problems for which exact solutions can be obtained via efficient algorithms!  
 
 However, in this paper, we show that in fact, \eqref{eq:nonconvex-nopos} is a rare $\ell_0$ optimization problem that can be \emph{exactly solved for the global optimum using an efficient algorithm}. 
  This is because \eqref{eq:nonconvex-nopos} can be seen as a changepoint detection problem, for which efficient algorithms that run in no more than $\mathcal{O}(T^2)$ time, and often closer to $\mathcal{O}(T)$ time, are available. 
  Furthermore, our implementation of the exact algorithm for solving \eqref{eq:nonconvex-nopos} yields excellent results relative to the convex approximation  \eqref{eq:convex-fried} considered by \cite{friedrich2016fast} and \cite{friedrichfast2017}. 
   This vastly improved performance can be seen in Figure~\ref{fig:overshrink}(d). 
          
   The rest of this paper is organized as follows. In Section~\ref{sec:algorithm}, we present an exact algorithm for solving the $\ell_0$ problem \eqref{eq:nonconvex-nopos}. 
   In Section~\ref{sec:simulation}, we investigate the performance of this algorithm, relative to the  algorithm of \cite{friedrich2016fast} and \cite{friedrichfast2017} for solving the $\ell_1$ problem \eqref{eq:convex-fried}, in a simulation study. In Section~\ref{sec:realdata}, we investigate the performances of both algorithms for spike train inference on a data set for which the true spike times are known 
   \citep{chen2013ultrasensitive, genie2015} and on a data set from the Allen Brain Observatory \citep{allenStimulusSet2016, hawrylycz2016inferring}. In Section~\ref{sec:extensions} we generalize the problem \eqref{eq:nonconvex-nopos} in order to allow for efficient estimation of an intercept term, and to accommodate an auto-regressive model of order $p>1$ \eqref{eq:model}. Finally, we close with a discussion in Section~\ref{sec:discussion}.  Technical details and additional  results can be found in the Appendix.


\section{An Exact Algorithm for Solving Problem \eqref{eq:nonconvex-nopos}}
\label{sec:algorithm}

 In Section~\ref{subsec:cp}, we show that problem \eqref{eq:nonconvex-nopos} can be viewed as a changepoint detection problem.
  In Sections~\ref{subsec:algT2} and \ref{subsec:algT}, we apply existing algorithms for changepoint detection in order to efficiently solve \eqref{eq:nonconvex-nopos} for the global optimum 
  in $\mathcal{O}(T^{2})$ and in substantially fewer than 
   $O(T^2)$ operations, respectively.  
  Timing results are presented in Section~\ref{subsec:timing}. We discuss selection of the tuning parameter $\lambda$ and auto-regressive parameter $\gamma$ in \eqref{eq:nonconvex-nopos}  in Appendix~\ref{subsec:cv}.

\subsection{Recasting  \eqref{eq:nonconvex-nopos} as a Changepoint Detection Problem} 
\label{subsec:cp}

Recall that our goal is to solve the $\ell_0$ optimization problem \eqref{eq:nonconvex-nopos}, or equivalently, to compute $\hat{c}_1,\ldots,\hat{c}_T$ that solve the optimization problem
$$\minimize{c_1,\ldots,c_T}{  \frac{1}{2} \sum_{t=1}^T \left( y_t - c_t \right)^2 + \lambda \sum_{t=2}^T 1_{\left( c_t - \gamma c_{t-1} \neq 0\right) }}.$$ 
 We estimate a spike event at the $t$th timestep if $\hat{c}_t \neq \gamma \hat{c}_{t-1}$. 
 (We refer to this as a ``spike event", rather than a spike, since $\hat{c}_t \neq \gamma \hat{c}_{t-1}$ indicates the presence of at least one spike at the $t$th timepoint, but 
  does not directly provide an estimate of the number of spikes.) 
 We now make two observations about this optimization problem.
\begin{enumerate}
 \item Given that a spike event is estimated at the $t$th timestep, the estimated calcium concentration at any time $t_1<t$ is  independent of the estimated calcium concentration at any time $t_2 \geq t$.
 \item Given that two spike events are estimated at the $t$th and $t'$th timesteps with $t<t'$, and no spike events are estimated in between the $t$th and $t'$th timesteps, the calcium concentration is estimated to decay exponentially between the $t$th and $t'$th timesteps.
\end{enumerate}
This motivates us to consider the relationship between \eqref{eq:nonconvex-nopos} and a \emph{changepoint detection problem} \citep{aue2013structural, braun1998statistical, davis2006structural, yao1988estimating, lee1995estimating, jackson2005algorithm, killick2012optimal, maidstone2016optimal}
 of the form 
 \begin{equation}
	\minimize{0= \tau_0 < \tau_1<\ldots<\tau_k < \tau_{k+1} = T,k}{\sum_{j=0}^{k}  
 \D(y_{(\tau_{j}+1):\tau_{j+1}}) + \lambda k },
 \label{eq:cp}
 \end{equation}
 where 
 \begin{equation}
 \label{eq:D}
 \D(y_{a:b}) \equiv \underset{c_{a},  c_t = \gamma c_{t-1}, \; t=a+1,\ldots,b}{\min}
  \left\{  \frac{1}{2} \sum_{t=a}^{b} \left( y_t - c_t \right)^2 \right\}.
  \end{equation}
   In \eqref{eq:cp},  we are simultaneously minimizing the objective over the times at which the changepoints ($\tau_1,\ldots,\tau_k$) occur and the number of changepoints ($k$); the parameter $\lambda$ controls the relative importance of these two terms.

   The following result establishes an equivalence between \eqref{eq:cp} and \eqref{eq:nonconvex-nopos}.
   \begin{proposition}
   There is a one-to-one correspondence between the set of estimated spike events in the solution to  \eqref{eq:nonconvex-nopos} and the set of changepoints $0=\tau_{0}, \tau_1,\ldots,\tau_k, \tau_{k+1} = T$ in the solution to \eqref{eq:cp}, in the sense that $\hat{c}_{t} \neq \gamma \hat{c}_{t-1}$ if and only if $t \in \{\tau_{1} + 1, \ldots, \tau_{k} + 1 \}$. 
       Furthermore, given the set of changepoints, the solution to \eqref{eq:nonconvex-nopos} takes the form
\begin{align*}
\hat{c}_t &= 
\begin{cases}
\gamma \hat{c}_{t-1} &   \tau_{j} + 2 \leq t \leq \tau_{j+1} \\
\frac{\sum_{t=\tau_{j} + 1}^{\tau_{j+1}} y_{t}\gamma^{t-(\tau_{j} + 1)}}{\sum_{t=\tau_{j}+1}^{\tau_{j+1}} \gamma^{2(t-(\tau_{j} + 1))}} & t = \tau_{j} + 1
\end{cases}, 
\end{align*}
for $j \in \left\{ 0,\ldots,k\right\}$.  
\label{prop:equiv}
\end{proposition}
   Proposition~\ref{prop:equiv} indicates that in order to solve \eqref{eq:nonconvex-nopos}, it suffices to solve \eqref{eq:cp}.
   (We note that due to a slight discrepancy between the conventions used in the changepoint detection literature and the notion of a spike event in this paper, the indexing in Proposition~\ref{prop:equiv} is a little bit awkward, in the sense that the $k$th spike event is estimated to occur at time $\tau_k+1$, rather than at time $\tau_k$.)

In the next two sections, we will make use of the following result.
\begin{proposition}
The quantity \eqref{eq:D} has a closed-form expression,  
\begin{align*}
\D(y_{a:b}) &= \sum_{t=a}^{b} \frac{y_{t}^{2}}{2}  - \C(y_{a:b}) \sum_{t=a}^{b} y_{t}\gamma^{t-a} + \frac{\C(y_{a:b})^{2}}{2}\sum_{t=a}^{b} \gamma^{2(t-a)}, \mbox{ where} \\ 
\C(y_{a:b}) &= \frac{\sum_{t=a}^{b} y_{t}\gamma^{t-a}}{\sum_{t=a}^{b} \gamma^{2(t-a)}}.
\end{align*}
Furthermore, given $\D(y_{a:b})$, we can calculate $\D(y_{a:(b+1)})$  in constant time.
\label{prop:segment-soln}
\end{proposition}
Propositions~\ref{prop:equiv} and \ref{prop:segment-soln} are proven in Appendix~\ref{sec:appendix}.

\subsection{An Algorithm for Solving \eqref{eq:nonconvex-nopos} in $\mathcal{O}(T^{2})$  Operations}
\label{subsec:algT2}

In this section, we apply a dynamic programming algorithm proposed by \citet{jackson2005algorithm} and \cite{auger1989algorithms} in order to solve the changepoint detection problem \eqref{eq:cp} for the global optimum in $\mathcal{O}(T^{2})$ time.  
   Due to the equivalence between \eqref{eq:cp} and \eqref{eq:nonconvex-nopos} established in Proposition~\ref{prop:equiv}, this algorithm also solves problem \eqref{eq:nonconvex-nopos}.

Roughly speaking, this algorithm  recasts the very difficult problem of choosing  the times of all  changepoints simultaneously into the much simpler problem of choosing the time of just the most recent  changepoint.  
 In greater detail, consider solving \eqref{eq:cp} on the first $s$ timesteps. Define  $F(0) \equiv -\lambda$, and for $s \geq 1$, define
\begin{align*}
F(s) &= 
 \underset{0= \tau_{0}< \tau_{1}< \cdots < \tau_{k} < \tau_{k+1} = s, k}{\mathrm{min}} \left\{ \sum_{j=0}^{k}  
 \D(y_{(\tau_{j}+1):\tau_{j+1}}) + \lambda k \right\} \\
&= 
\underset{0= \tau_{0}< \tau_{1}< \cdots < \tau_{k} < \tau_{k+1} = s, k}{\mathrm{min}}\left\{ \sum_{j=0}^{k} \left[\D(y_{(\tau_{j} + 1):\tau_{j+1}}) + \lambda \right] - \lambda \right\}\\
&= \underset{0= \tau_{0}< \tau_{1}< \cdots < \tau_{k} < \tau_{k+1} = s, k}{\mathrm{min}}\left\{ \sum_{j=0}^{k-1} \left[\D(y_{(\tau_{j} + 1):\tau_{j+1}}) + \lambda \right] - \lambda + \D(y_{(\tau_{k} + 1):\tau_{k+1}}) + \lambda  \right\} \\
&= \underset{0\leq \tau_{k} < \tau_{k+1} = s}{\mathrm{min}} \left\{ \underset{0= \tau_{0}< \tau_{1}< \cdots < \tau_{k'} < \tau_{k'+1} = \tau_{k}, k'}{\mathrm{min}} \left\{ \sum_{j=0}^{k'} \left[\D(y_{(\tau_{j} + 1):\tau_{j+1}}) + \lambda \right] - \lambda \right\}+ \D(y_{(\tau_{k} + 1):\tau_{k+1}}) + \lambda\right\} \\
&= \underset{0\leq \tau < s}{\mathrm{min}}\left\{ F(\tau) + \D(y_{(\tau + 1):s}) + \lambda \right\}. \numberthis \label{eq:op-recursion}
\end{align*}
In other words, in order to solve  \eqref{eq:cp}, we need simply identify the time of the most recent changepoint, and then solve  \eqref{eq:cp} on all earlier times.

This recursion gives a simple recipe for evaluating $F(T)$ efficiently: set $F(0) = -\lambda$, and compute $F(1), F(2), \ldots, F(T)$ based on previously calculated (and stored) values. For example, at $s = 1$, calculate and store 
\begin{align*}
F(1) = \underset{0\leq \tau < 1}{\mathrm{min}} \left\{ F(\tau) + \D(y_{(\tau + 1):1}) + \lambda \right\} = F(0) + \D(y_{1}) + \lambda, 
\end{align*}
and then at $s=2$ use the previously calculated values $F(0)$ and $F(1)$ to compute the minimum over a finite set with $s$ elements
\begin{align*}
F(2) = \underset{\tau \in \{0, 1\}}{\mathrm{min}} \left\{ F(\tau) + \D(y_{(\tau + 1):2}) + \lambda \right\} 
= \min\left\{F(0) + \D(y_{1:2}) + \lambda, F(1) + \D(y_{2}) + \lambda\right\}. \end{align*}
Given $F(1), \ldots, F(s-1)$, computing $F(s)$ requires minimizing over a finite set of size $s$, and therefore it has computational cost linear in $s$. The total  cost of computing $F(T)$ is quadratic in the total number of timesteps, $T$, since there are $T+1$ subproblems: $\sum_{s=0}^{T} s =  \mathcal{O}(T^{2})$.

Full details are provided in Algorithm~\ref{alg:op}. We note that this algorithm is particularly efficient in light of Proposition~\ref{prop:segment-soln}, which makes it possible to perform a constant-time update to $\D(y_{(\tau+1):s})$ in order to compute $\D(y_{(\tau+1):(s+1)})$. 
\begin{algorithm}[h!]
    \SetKwInOut{Initialize}{Initialize}
    \SetKwInOut{Output}{Output}
	\Initialize{$F(0) = -\lambda$, $cp(0) = \emptyset$}
	\ForEach{$s = 1, 2, \ldots, T$}{
		Calculate $F(s) = \minn{0 \leq \tau < s}{ F(\tau) + \D(y_{(\tau+1):s}) + \lambda}$ \\
		Set $s' = \argmin{0 \leq \tau < s}{ F(\tau) + \D(y_{(\tau+1):s}) + \lambda}$ \\
		Update $cp(s) = (cp(s'), s')$
	}
    \Output{The number of spike events $k\equiv \card(cp(T))$, the changepoints $\{\tau_1,\ldots,\tau_k\} \equiv cp(T)$, the spike times $\{ \tau_{1} +1, \ldots, \tau_{k} + 1\}$, and the estimated calcium concentrations
\begin{align*}
\hat{c}_t &\equiv
\begin{cases}
\gamma \hat{c}_{t-1} &  \tau_{j} + 2 \leq t \leq \tau_{j+1} \\
\frac{\sum_{t=\tau_{j} + 1}^{\tau_{j+1}} y_{t}\gamma^{t-(\tau_{j} + 1)}}{\sum_{t=\tau_{j}+1}^{\tau_{j+1}} \gamma^{2(t-(\tau_{j} + 1))}} & t = \tau_{j} + 1
\end{cases}, 
\end{align*} for $j = 0, \ldots, k$, where $\tau_{0} = 0$.} 
        \caption{An $\mathcal{O}(T^{2})$ Algorithm for Solving \eqref{eq:nonconvex-nopos}}\label{alg:op}
\end{algorithm}



\subsection{Dramatic Speed-Ups Using Cost-Complexity Pruning}

\label{subsec:algT}

In a recent paper, \citet{killick2012optimal} 
considered  problems of the form \eqref{eq:cp} for which an assumption on $\D(\cdot)$ holds; this assumption is satisfied by \eqref{eq:D}.

The main insight of their paper is as follows. Suppose that $s < r$ and $F(s) + \D(y_{(s+1):r}) > F(r)$. Then for any $q>r$, it is mathematically impossible for the most recent changepoint before the $q$th timestep to have occurred at the $s$th timestep.  
 This allows us to \emph{prune} the set  of candidate changepoints that must be considered in each step of Algorithm~\ref{alg:pelt}, leading to drastic speed-ups. Details are provided in Algorithm~\ref{alg:pelt}, which solves \eqref{eq:nonconvex-nopos} for the global optimum. 
 
Under several technical assumptions, \citet{killick2012optimal} show that the expected complexity of this algorithm is $\mathcal{O}(T)$. The main assumption is that the expected number of changepoints in the data increases linearly with the length of the data; this is reasonable in the context of calcium imaging data, in which we expect the number of neuron spike events to be linear in the length of the recording. 

\begin{algorithm}[h!]
    \SetKwInOut{Initialize}{Initialize}
    \SetKwInOut{Output}{Output}
	\Initialize{$F(0) = -\lambda$, $cp(0) = \emptyset$, $\ES_{1} = \{0\}$}
	\ForEach{$s = 1, 2, \ldots, T$}{
		Calculate $F(s) = \minn{\tau \in \ES_{s}}{ F(\tau) + \D(y_{(\tau+1):s}) + \lambda}$ \\
		Set $s' = \argmin{\tau \in \ES_{s}}{ F(\tau) + \D(y_{(\tau+1):s}) + \lambda}$  \\
		Update $\ES_{s + 1} = \{ \tau \in \{ \ES_{s} \cup s\} : F(\tau) + \D(y_{(\tau+1):s}) < F(s)\}$  \\
		Update $cp(s) = (cp(s'), s')$
	}
    \Output{The number of spike events $k\equiv \card(cp(T))$, the changepoints $\{\tau_1,\ldots,\tau_k\} \equiv cp(T)$, the spike times $\{ \tau_{1} +1, \ldots, \tau_{k} + 1 \}$, and the estimated calcium concentrations
\begin{align*}
\hat{c}_t &\equiv
\begin{cases}
\gamma \hat{c}_{t-1} &  \tau_{j} + 2 \leq t \leq \tau_{j+1} \\
\frac{\sum_{t=\tau_{j} + 1}^{\tau_{j+1}} y_{t}\gamma^{t-(\tau_{j} + 1)}}{\sum_{t=\tau_{j}+1}^{\tau_{j+1}} \gamma^{2(t-(\tau_{j} + 1))}} & t = \tau_{j} + 1
\end{cases}, 
\end{align*} for $j = 0, \ldots, k$, where $\tau_{0} = 0$.} 
\caption{An Algorithm for Solving \eqref{eq:nonconvex-nopos} in Substantially Fewer than $\mathcal{O}(T^2)$ Operations}\label{alg:pelt}
\end{algorithm}

\subsection{Timing Results for Solving \eqref{eq:nonconvex-nopos}}
\label{subsec:timing}

 We simulated data from \eqref{eq:model} with $\gamma= 0.998$, $\sigma = 0.15$, and  $s_t \sim_\text{ind.} \text{Poisson}(\theta)$ with  $\theta\in\{ 0.1, 0.01, 0.001\}$. 
 We solved \eqref{eq:nonconvex-nopos} with  $\lambda = 1$, using our \verb=R=-language implementations of Algorithms~\ref{alg:op} and \ref{alg:pelt}.

\begin{figure}[h!]
\begin{center}
\includegraphics[scale = 0.6]{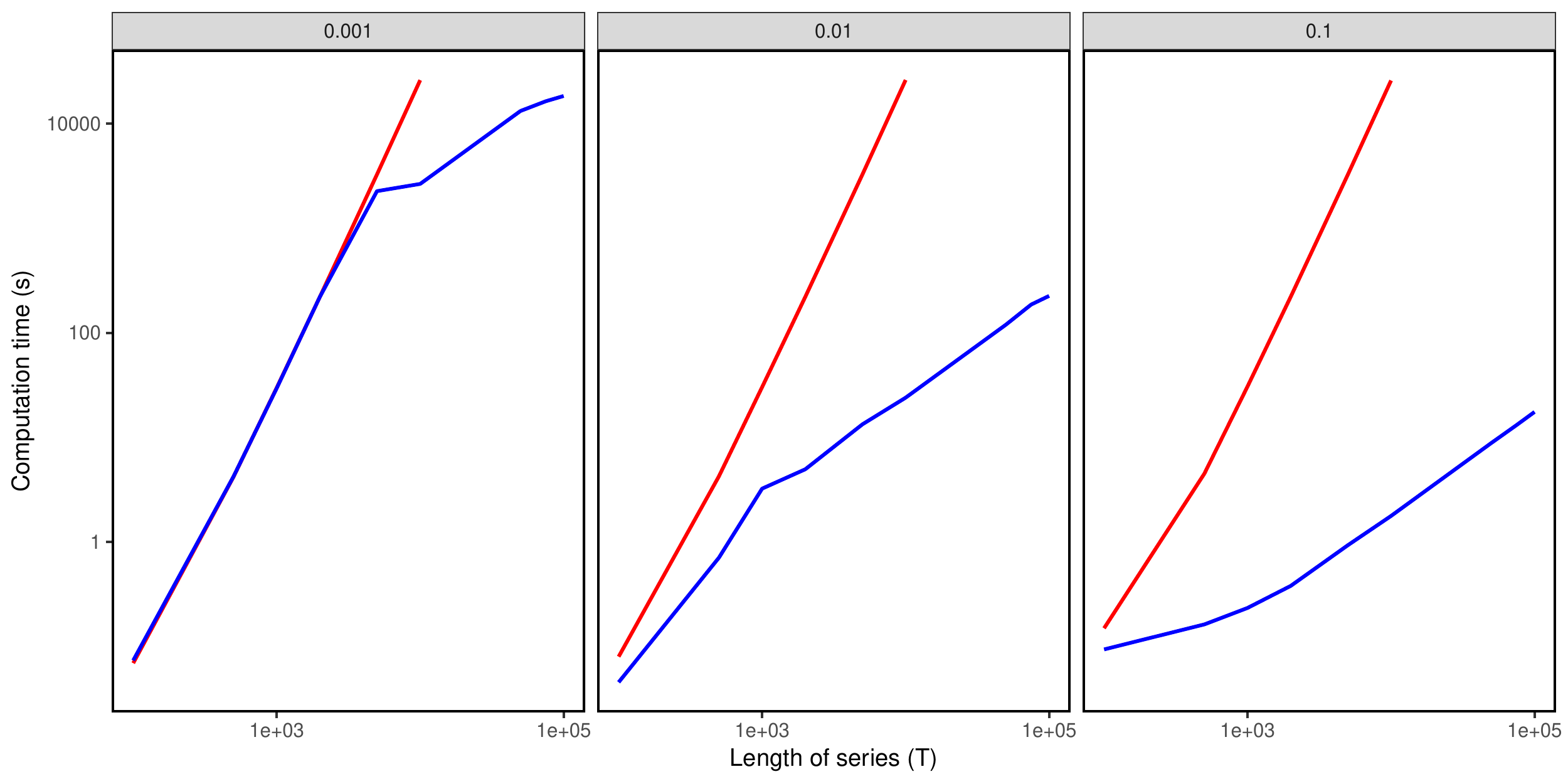}
\caption{Timing results   for solving \eqref{eq:nonconvex-nopos} for the global optimum, using Algorithms~\ref{alg:op} (\protect\includegraphics[height=0.4em]{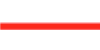}) and \ref{alg:pelt} (\protect\includegraphics[height=0.4em]{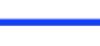}). The $x$-axis displays the length of the time series ($T$), and the $y$-axis displays the average running time, in seconds. Each panel, from left to right, corresponds to data simulated according to \eqref{eq:model} with $s_t \sim_\text{i.i.d.} \text{Poisson}(\theta)$, with $\theta \in \{0.001, 0.01, 0.1\}$. Standard errors are on average $< 0.1\%$ of the mean compute time. Additional details are provided in Section~\ref{subsec:timing}.}
 \label{fig:timing}
\end{center}
\end{figure}

 Timing results, averaged over  50 simulated data sets, are displayed in 
Figure~\ref{fig:timing}.  
 As expected, the running time of Algorithm~\ref{alg:op} scales quadratically in the length of the time series, whereas the running time of 
  Algorithm~\ref{alg:pelt} is upper bounded by that of Algorithm~\ref{alg:op}. Furthermore, the running time of Algorithm~\ref{alg:pelt} decreases as the firing rate increases. The \cite{chen2013ultrasensitive} dataset explored in Section~\ref{subsec:chen} has firing rate on the same order of magnitude as the middle panel, $\theta = 0.01$. 
 Using Algorithm~\ref{alg:pelt}, we can solve \eqref{eq:nonconvex-nopos} for the global optimum in a few minutes  on a 2.5 GHz Intel Core i7 Macbook Pro on fluorescence traces of length $100,000$  with moderate to high firing rates. 

We note here that Algorithm~\ref{alg:pelt} for solving \eqref{eq:nonconvex-nopos} is much slower than the algorithm of \citet{friedrichfast2017} for solving \eqref{eq:convex-fried}, which is implemented in Cython and has approximately linear running time. It should be possible to develop a faster algorithm for solving \eqref{eq:nonconvex-nopos} using ideas from \cite{johnson2013dynamic}, \cite{maidstone2016optimal}, and \cite{hocking2017log}. 
Furthermore, a much faster implementation of Algorithm~\ref{alg:pelt} would be possible using a language other than \verb=R=. We leave such improvements to future work.



\section{Simulation Study}
\label{sec:simulation}

\subsection{Comparison Methods} 
\label{sec:comparison-methods}

In this section, we use  \emph{in silico} data  to demonstrate the performance advantages of the $\ell_{0}$  approach \eqref{eq:nonconvex-nopos} over two competing approaches: 
\begin{enumerate}
\item The $\ell_{1}$ proposal \eqref{eq:convex-fried} of \citet{friedrich2016fast} and \citet{friedrichfast2017}, which involves a single tuning parameter $\lambda$.
\item A thresholded version of  the $\ell_1$ estimator. Letting $\hat{s}_{2}, \ldots, \hat{s}_{T}$ denote the solution to \eqref{eq:convex-fried},  we define the \emph{post-thresholding estimator} as   
\begin{equation}
\tilde{s}_{t} = \hat{s}_{t} 1_{(\hat{s}_{t} \geq L)}, \quad t = 2, \ldots, T,
\label{eq:post}
\end{equation}
 for $L$ a positive constant. 
 In other words, the post-thresholding estimator retains only the estimated spikes for which the estimated increase in calcium exceeds a threshold $L$. 
 The post-thresholding estimator involves two tuning parameters: $\lambda$ in \eqref{eq:convex-fried}, as well as the value of $L$ used to perform thresholding.  
\end{enumerate}
The post-thresholding estimator is motivated by the fact that the solution to \eqref{eq:convex-fried} tends to yield many ``small" spikes: i.e. $\hat{s}_t$ is near zero, but not exactly equal to zero, for many timesteps. In fact, this can be seen in 
 Figure~\ref{fig:overshrink}(c).
  As seen in Figure~\ref{fig:overshrink}(e), the post-thresholding estimator has the potential to improve the performance of the $\ell_1$ estimator by removing some of these small spikes. Of course, the post-thresholding estimator with $L=0$ is identical to the $\ell_1$ estimator from \eqref{eq:convex-fried}. 

\subsection{Performance Measures}

We measure performance of each method based on two criteria: (i) error in calcium estimation, and (ii) error in spike detection.

We consider the mean of squared differences between the true calcium concentration in \eqref{eq:model} and the estimated calcium concentration that solves \eqref{eq:nonconvex-nopos},
\begin{equation}
\mathrm{MSE}(c, \hat{c}) = \frac1T \sum_{t = 1}^{T} (c_{t} - \hat{c}_{t})^{2}.
\label{eq:calciumerror}
\end{equation}
This quantity involves the unobserved calcium concentrations, $c_1,\ldots,c_T$, and thus can only be computed on simulated data. Furthermore, this quantity can be computed for our $\ell_0$ proposal \eqref{eq:nonconvex-nopos} and for the $\ell_1$ proposal \eqref{eq:convex-fried}, but not for the post-thresholding estimator \eqref{eq:post}, since the post-thresholding estimator does not lead to an estimate of the underlying calcium concentrations.

We now consider the task of quantifying the error in spike detection.  
 We make use of the Victor-Purpura distance metric \citep{victor1996nature, victor1997metric}, which defines the distance between two spike trains as the minimum cost of transforming one spike train to the other through spike insertion, deletion, or translation. We also  use  the van Rossum distance \citep{van2001novel}, 
defined as the mean squared difference between two spike trains that have been convolved with an exponential kernel with timescale $\tau = 2$. 

\subsection{Results}

We generated 100 simulated data sets according to  \eqref{eq:model} with parameter settings $\gamma = 0.96$, $T = 5000$, $\sigma = 0.15$, and $s_t \sim_\text{i.i.d.} \text{Poisson(0.01)}$.


On each simulated data set, we solved \eqref{eq:nonconvex-nopos} and \eqref{eq:convex-fried} for a range of values of the tuning parameter $\lambda$. Moreover, we post-thresholded the $\ell_{1}$ solution, as in  \eqref{eq:post}, with five different threshold values:  $L\in\{0, 0.125, 0.250, 0.375, 0.500 \}$.

\begin{figure}[h]
\begin{center}
\hspace{5mm} (a) \hspace{50mm} (b) \hspace{50mm} (c) 
\includegraphics[scale = 0.42]{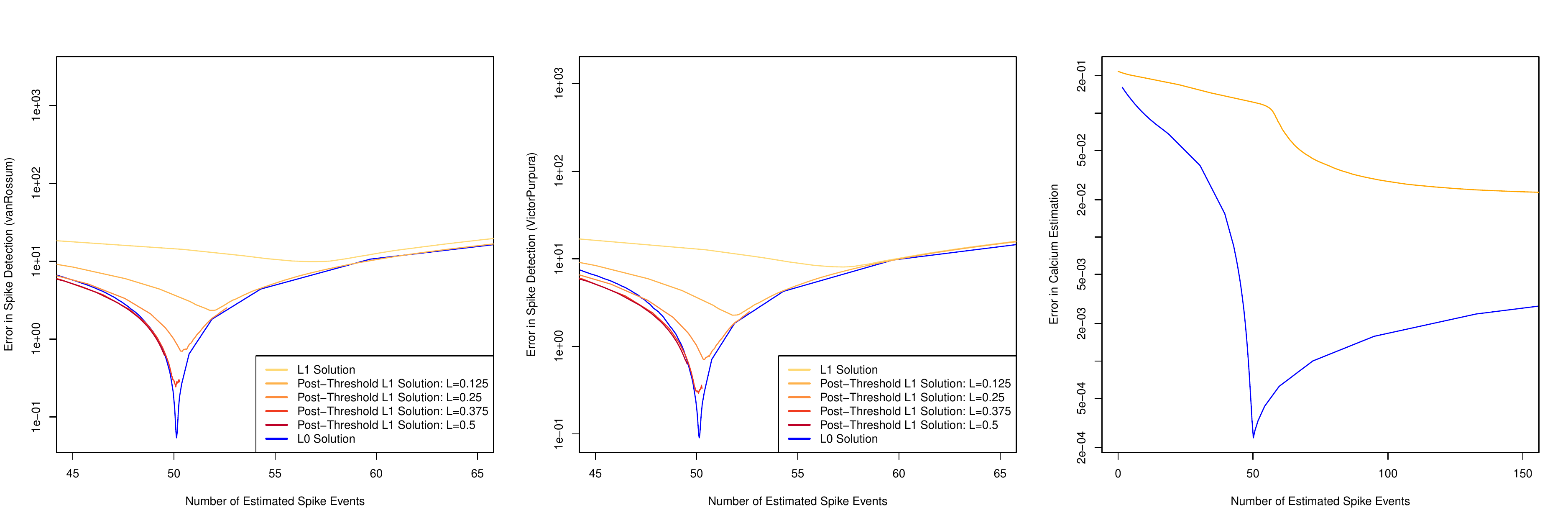}
\caption{Simulation study to assess the error in spike detection and calcium estimation, for the $\ell_1$ \eqref{eq:convex-fried}, post-thresholded $\ell_{1}$ \eqref{eq:post}, and $\ell_0$ \eqref{eq:nonconvex-pis1} problems.
\emph{(a)}: Error in spike detection, measured using van Rossum distance. \emph{(b)}: Error in spike detection, measured using Victor-Purpura distance. \emph{(c)}: Error in calcium estimation \eqref{eq:calciumerror}.  
 Simulation details are provided in Section~\ref{sec:simulation}.}
\label{fig:mse}
\end{center}
\end{figure}

  
Figure~\ref{fig:mse}(a) displays the error in spike event detection for the van Rossum distance, Figure~\ref{fig:mse}(b) displays the error in spike event detection for the Victor-Purpura distance metric, and Figure~\ref{fig:mse}(c) displays the error in calcium estimation \eqref{eq:calciumerror}, for the $\ell_0$ problem \eqref{eq:nonconvex-nopos} and the $\ell_1$ problem \eqref{eq:convex-fried}, for a range of values of $\lambda$.  Results are averaged over the 50 simulated data sets. 

As mentioned earlier, since the calcium concentration is not defined for the post-thresholding estimator \eqref{eq:post}, the post-thresholding estimator is not displayed in Figure~\ref{fig:mse}(c).
 In Figures~\ref{fig:mse}(a) and \ref{fig:mse}(b),   five distinct curves are displayed for the post-thresholding operator; each corresponds to a distinct value of $L$. 
 Note 
that as $L$ increases, the maximum possible  number of estimated spikes from the post-thresholding estimator decreases.   For example, with $\lambda=0$ and $L=0.5$, no more than approximately 50 spikes are estimated by the post-thresholding estimator. For this reason, some of the curves corresponding to the post-thresholding estimator appear truncated in Figures~\ref{fig:mse}(a) and \ref{fig:mse}(b).  

Figure~\ref{fig:mse} reveals that the $\ell_0$ estimator \eqref{eq:nonconvex-nopos} results in dramatically lower errors in both calcium estimation and spike detection than  the  $\ell_1$ estimator \eqref{eq:convex-fried} (which is equivalent to the post-thresholding operator with $L=0$). Although post-thresholding with $L>0$ improves upon the unthresholded $\ell_{1}$ estimator, the $\ell_{0}$ estimator still substantially outperforms all competitors in Figures~\ref{fig:mse}(a) and \ref{fig:mse}(b). Moreover, the $\ell_0$ estimator requires just a single tuning parameter $\lambda$ in \eqref{eq:nonconvex-nopos}, whereas the post-thresholding procedure involves two tuning parameters, $\lambda$ in \eqref{eq:convex-fried} and $L$ in \eqref{eq:post}, leading to challenges in tuning parameter selection.

Furthermore, the $\ell_0$ problem \eqref{eq:nonconvex-nopos} achieves the lowest errors in both calcium estimation and spike detection when applied using a value of the tuning parameter $\lambda$ that yields approximately 50 estimated spikes, which is the expected number of spikes in this simulation. This suggests that it should be possible to use a cross-validation scheme to select the tuning parameter $\lambda$ for the $\ell_0$ approach; we propose such a scheme in Appendix~\ref{subsec:cv}. By contrast, in Figure~\ref{fig:mse}(b), the $\ell_1$ approach achieves its lowest error in calcium estimation when far more than 50 spikes are estimated. This is a consequence of the fact that the $\ell_1$ penalty simultaneously reduces the number of estimated spikes and shrinks the estimated calcium. Therefore, the value of the tuning parameter $\lambda$ in \eqref{eq:convex-fried} that yields the most accurate estimate of calcium will result in severe over-estimation of the number of spikes. This means that the cross-validation scheme detailed in Appendix~\ref{subsec:cv} will not perform well for the $\ell_1$ approach.

\section{Application to Calcium Imaging Data}
\label{sec:realdata}

In this section, we apply our $\ell_0$ proposal \eqref{eq:nonconvex-nopos} and the $\ell_1$ proposal  of \cite{friedrich2016fast} and \cite{friedrichfast2017} \eqref{eq:convex-fried}, both with and without post-thresholding \eqref{eq:post}, to two calcium imaging data sets. 
 In the first data set, the true spike times are known \citep{chen2013ultrasensitive, genie2015}, and so we can directly assess the spike detection accuracy of each proposal. 
  In the second data set, the true spike times are unknown \citep{allenStimulusSet2016, hawrylycz2016inferring}; nonetheless, we are able to make a qualitative comparison of the results of the $\ell_1$ and $\ell_0$ proposals.

\subsection{Application to \cite{chen2013ultrasensitive} Data}
\label{subsec:chen}

We first consider a data set that consists of 
simultaneous calcium imaging and electrophysiological measurements  \citep{chen2013ultrasensitive,genie2015}, obtained from the Collaborative Research in Computational Neuroscience portal (\url{http://crcns.org/data-sets/methods/cai-1/about-cai-1}). In what follows, we refer to the spike times inferred from the electrophysiological measurements  as the ``true" spikes.

\begin{figure}[htbp]
\begin{center}
\includegraphics[scale = 0.5]{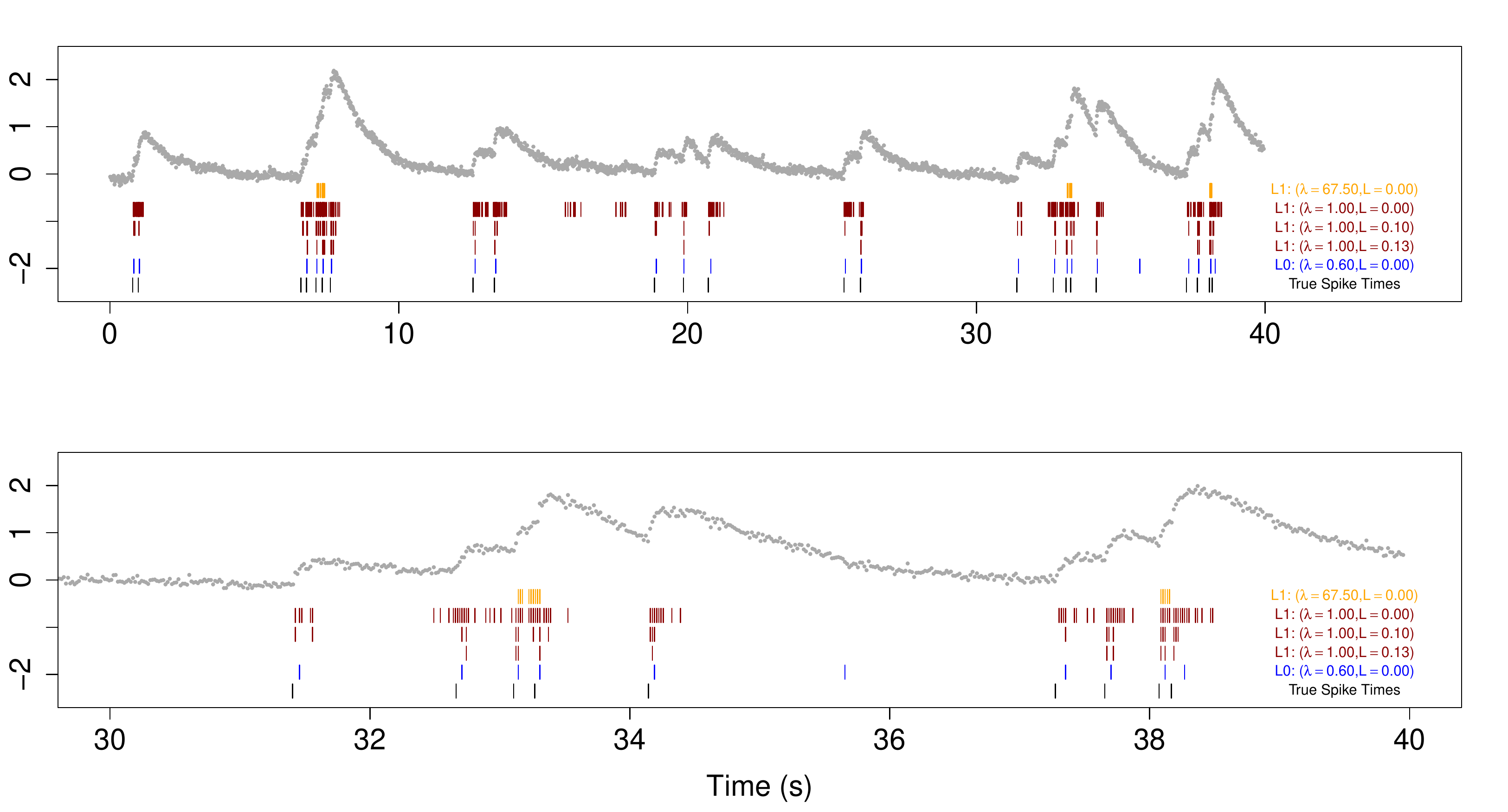}
\caption{Spike detection for  cell 2002 of the \citet{chen2013ultrasensitive} data. The observed fluorescence (\protect\includegraphics[height=0.4em]{grey-circle}) and true spikes (\protect\includegraphics[height=0.3em]{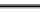}) are displayed. Estimated  spike times from the $\ell_{0}$ problem \eqref{eq:nonconvex-pis1} are shown in (\protect\includegraphics[height=0.4em]{blue}),  estimated spike times from the $\ell_{1}$ problem \eqref{eq:convex-fried} are shown in (\protect\includegraphics[height=0.4em]{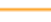}), and estimated spike times from the post-thresholding estimator \eqref{eq:post} are shown in (\protect\includegraphics[height=0.4em]{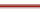}). Times $0s-35s$ are shown in the top row; the second row zooms into time $30s-40s$ in order to illustrate the behavior around a large increase in calcium concentration.}
\label{fig:chen1}
\end{center}
\end{figure}

The top panel of Figure~\ref{fig:chen1} shows a 40-second recording from cell 2002, which expresses GCaMP6s. The data are measured at 60 Hz, for a total of 2400 timesteps. The raw fluorescence traces are DF/F transformed  with a $20\%$ percentile filter as in Figure~3 of \citet{friedrichfast2017}.
In this 40-second recording, there are a total of 23 true spikes; therefore,  we solved the $\ell_0$ and $\ell_1$ problems with $\gamma = 0.9864405$ using values of $\lambda$ in \eqref{eq:nonconvex-nopos} and \eqref{eq:convex-fried} that yield 23 estimated spikes. Additionally, we solved the $\ell_{1}$ problem with $\lambda = 1$, and post-thresholded it according to \eqref{eq:post} using $L=0$, $0.1$, and $0.13$; these threshold values yielded   $230$, $54$, and $23$ estimated spikes, respectively.

Figure~\ref{fig:chen1} displays the estimated spikes resulting from the $\ell_0$ proposal, the estimated spikes resulting from the $\ell_1$ proposal, the estimated spikes from post-thresholding the $\ell_{1}$ solutions, and the ground truth spikes. 
  We see that the  $\ell_0$ proposal has one false negative (i.e. it misses one true spike at around $7$ seconds) 
 and one false positive (i.e. it estimates a spike at around 36 seconds, 
 where there is no true spike). 
 By contrast, the $\ell_1$ problem concentrates the 23 estimated spikes at three points in time, and therefore suffers from a substantial number of false positives as well as false negatives.
 Because the $\ell_1$ penalty controls both the number of spikes and the estimated calcium, the $\ell_1$ problem tends to put a large number of spikes in a row, each of which is associated with  a very modest increase in calcium. This is consistent with the results seen in Figures~\ref{fig:overshrink} and \ref{fig:mse}. Post-thresholding the $\ell_1$ estimator does lead to an improvement in results relative to the unthresholded $\ell_1$ method; however,  the post-thresholded solution with 23 spikes still tends to estimate a number of spikes in short succession when in fact only one true spike is present, and also misses several true spike events.
 
We note that the $\ell_{0}$ method tends to estimate spike times one or two timesteps ahead of the true spike times. This is due to model misspecification: model \eqref{eq:model} with $p=1$ assumes that the calcium concentration increases instantaneously due to a spike event, and subsequently decays; however, we see from Figure~\ref{fig:chen1}  that in reality, a spike event is followed by an increase in calcium over the course of  a few timesteps, before the onset of  exponential decay. We see two possible avenues to address this relatively minor issue: estimated spike times from the $\ell_{0}$ method can be adjusted to account for this empirical observation; or else the optimization problem \eqref{eq:nonconvex-nopos} can be adjusted in order  to allow for more realistic calcium dynamics (for example, by solving an $\ell_0$ optimization problem corresponding to \eqref{eq:model} with $p>1$). We explore the second alternative in Section~\ref{sec:extensions}. 
 
In Appendix~\ref{sec:heur}, we apply an approach proposed by \citet{friedrichfast2017} to approximate the solution to a non-convex problem using a greedy algorithm. This alternative approach performs quite a bit better than solving the $\ell_{1}$ problem \eqref{eq:convex-fried};  however, it does not achieve the global optimum.

\subsection{Application to Allen Brain Observatory Data}

We now consider a data set from the Allen Brain Observatory, a large open-source repository of calcium imaging data from the mouse visual cortex \citep{allenStimulusSet2016, hawrylycz2016inferring}. For this data, the true spike times are not available, and so it is difficult to objectively assess the performances of the $\ell_1$,  post-thresholded $\ell_1$, and $\ell_0$ methods. Instead, for each method we present several fits that differ in the number of detected spikes.   We argue that the  $\ell_{0}$ problem yields results that are qualitatively superior to those of the competitors, in the sense that they are better supported by  visual inspection of the data.

\begin{landscape}
\begin{figure}
\centering
\begin{subfigure}[b]{0.4\textheight}
\centering
\includegraphics[scale=.5]{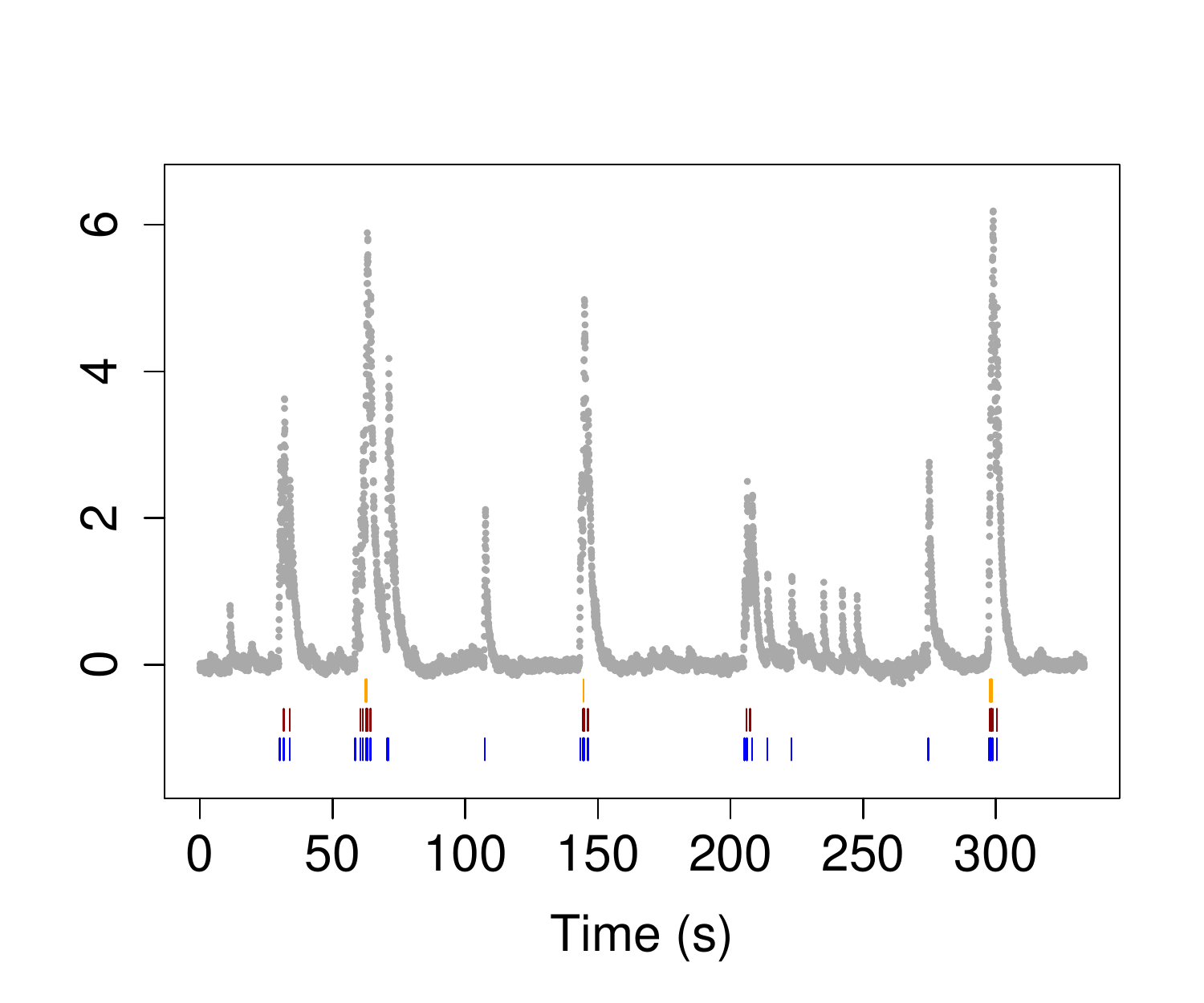} 
\caption{27 spikes.}
\end{subfigure}
\begin{subfigure}[b]{0.4\textheight}
\centering
\includegraphics[scale=.5]{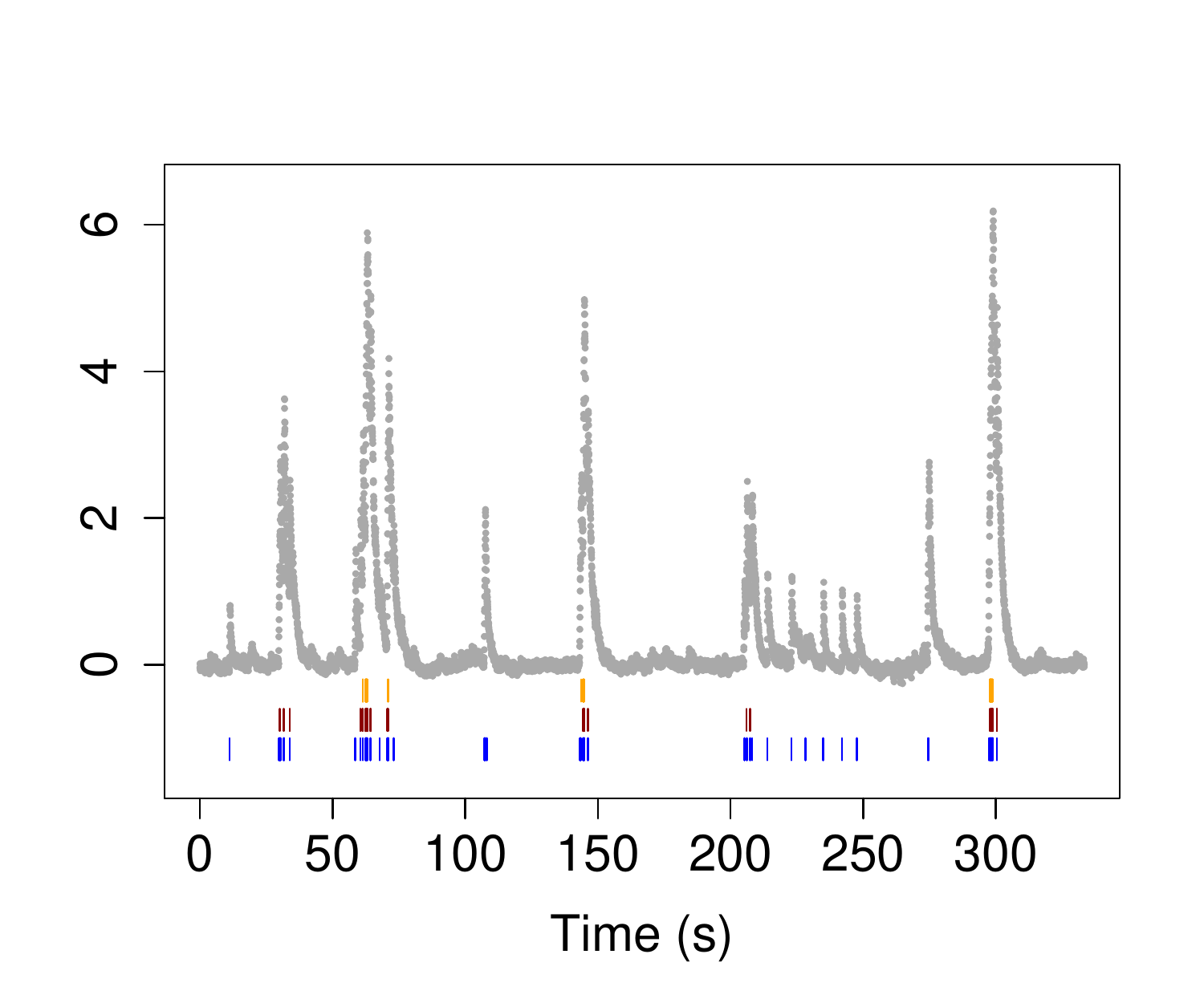} 
\caption{49 spikes.}
\end{subfigure}
\begin{subfigure}[b]{0.4\textheight}
\centering
\includegraphics[scale=.5]{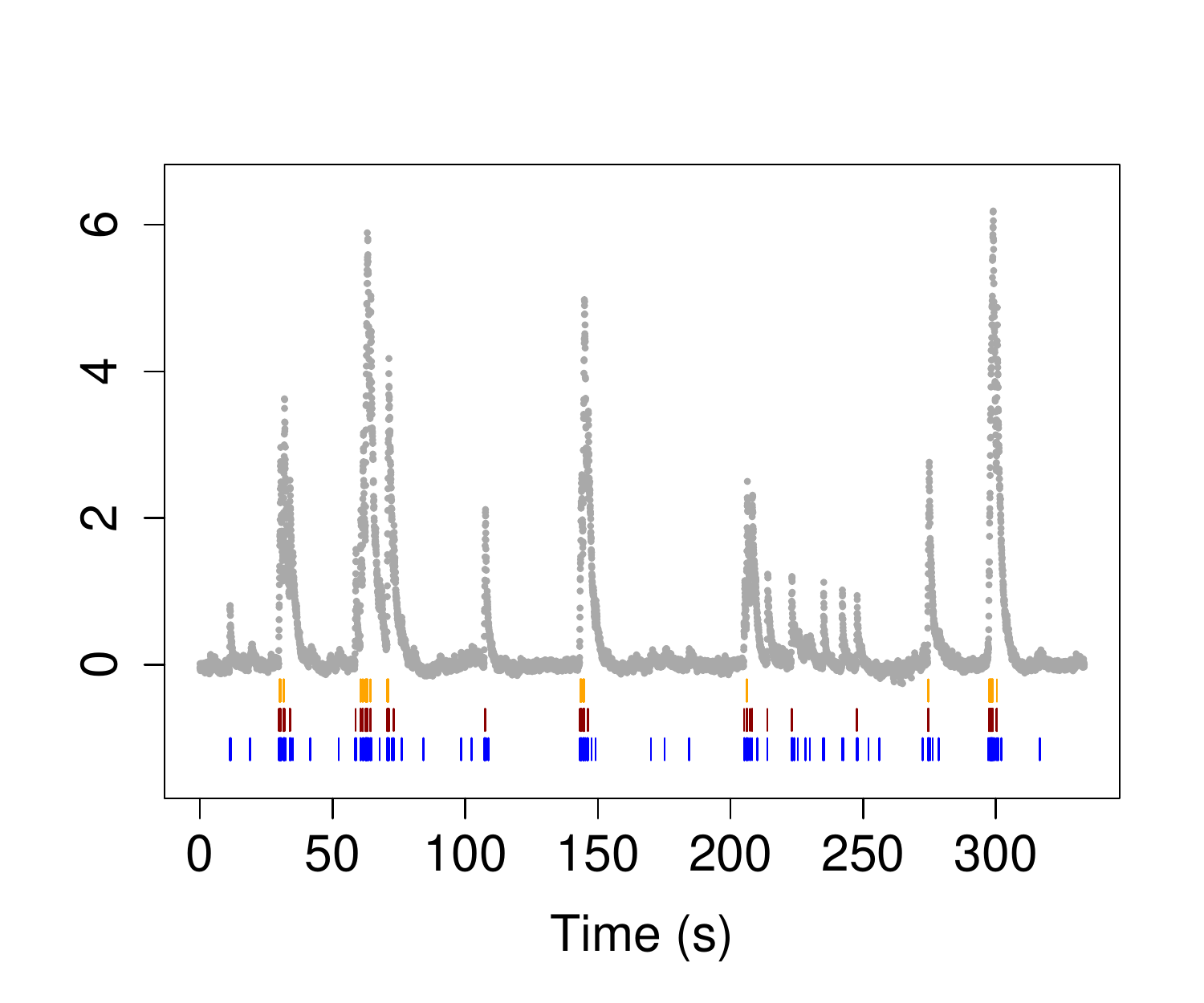} 
\caption{128 spikes.}
\end{subfigure}%

\begin{subfigure}[b]{0.4\textheight}
\centering
\includegraphics[scale=.5]{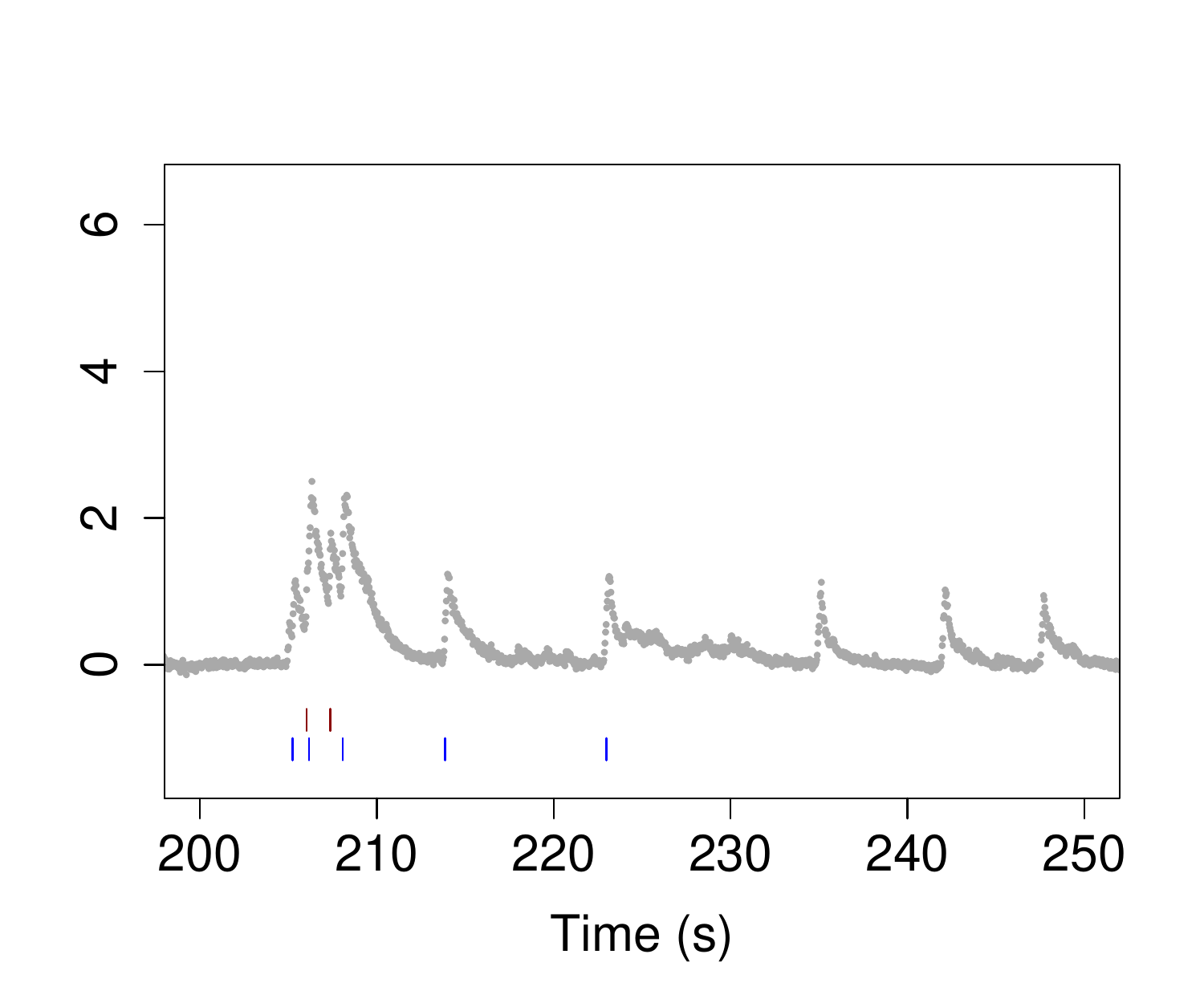}
\caption{Zoomed 27 spikes.}
\end{subfigure}
\begin{subfigure}[b]{0.4\textheight}
\centering
\includegraphics[scale=.5]{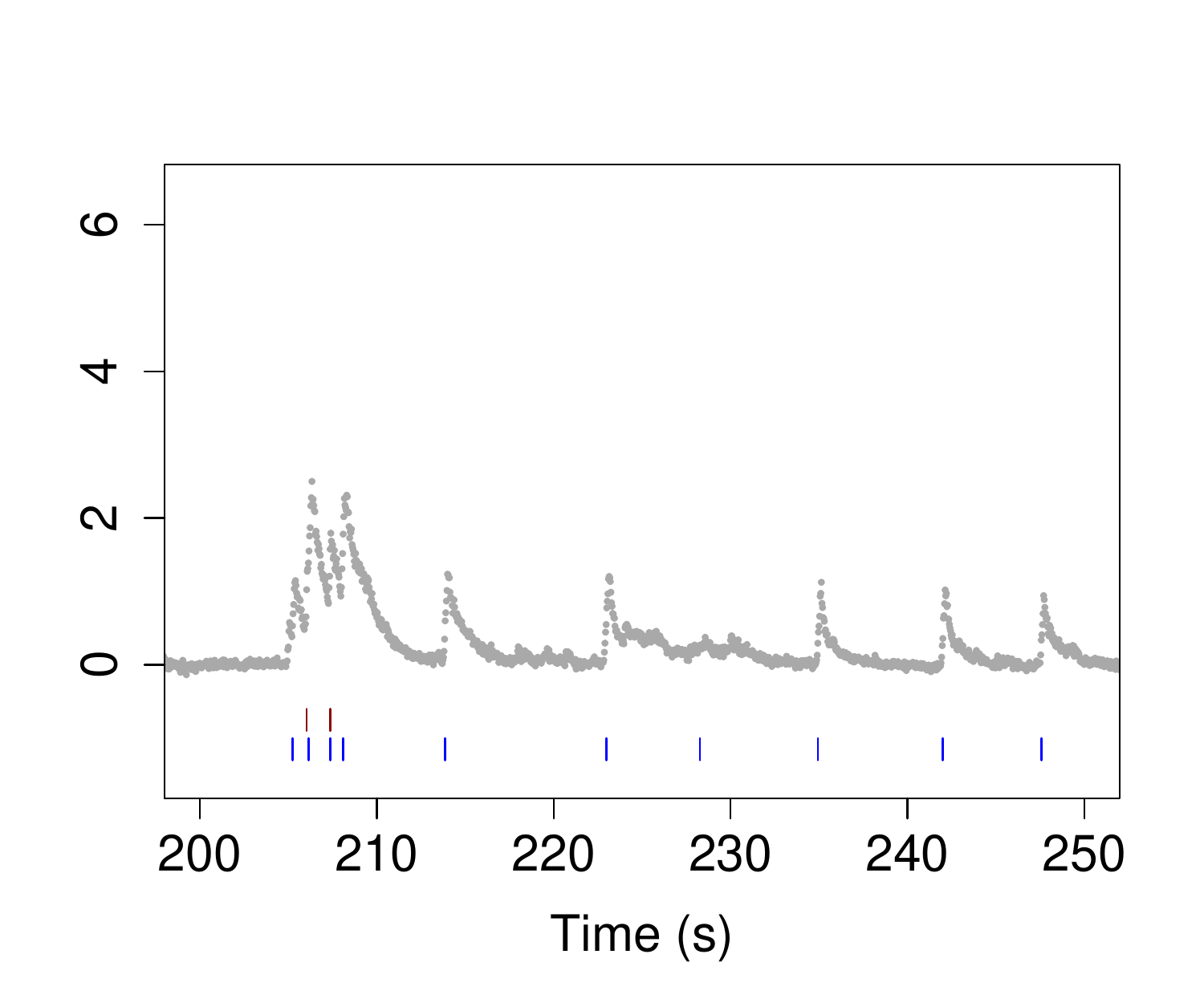} 
\caption{Zoomed 49 spikes.}
\end{subfigure}
\begin{subfigure}[b]{0.4\textheight}
\centering
\includegraphics[scale=.5]{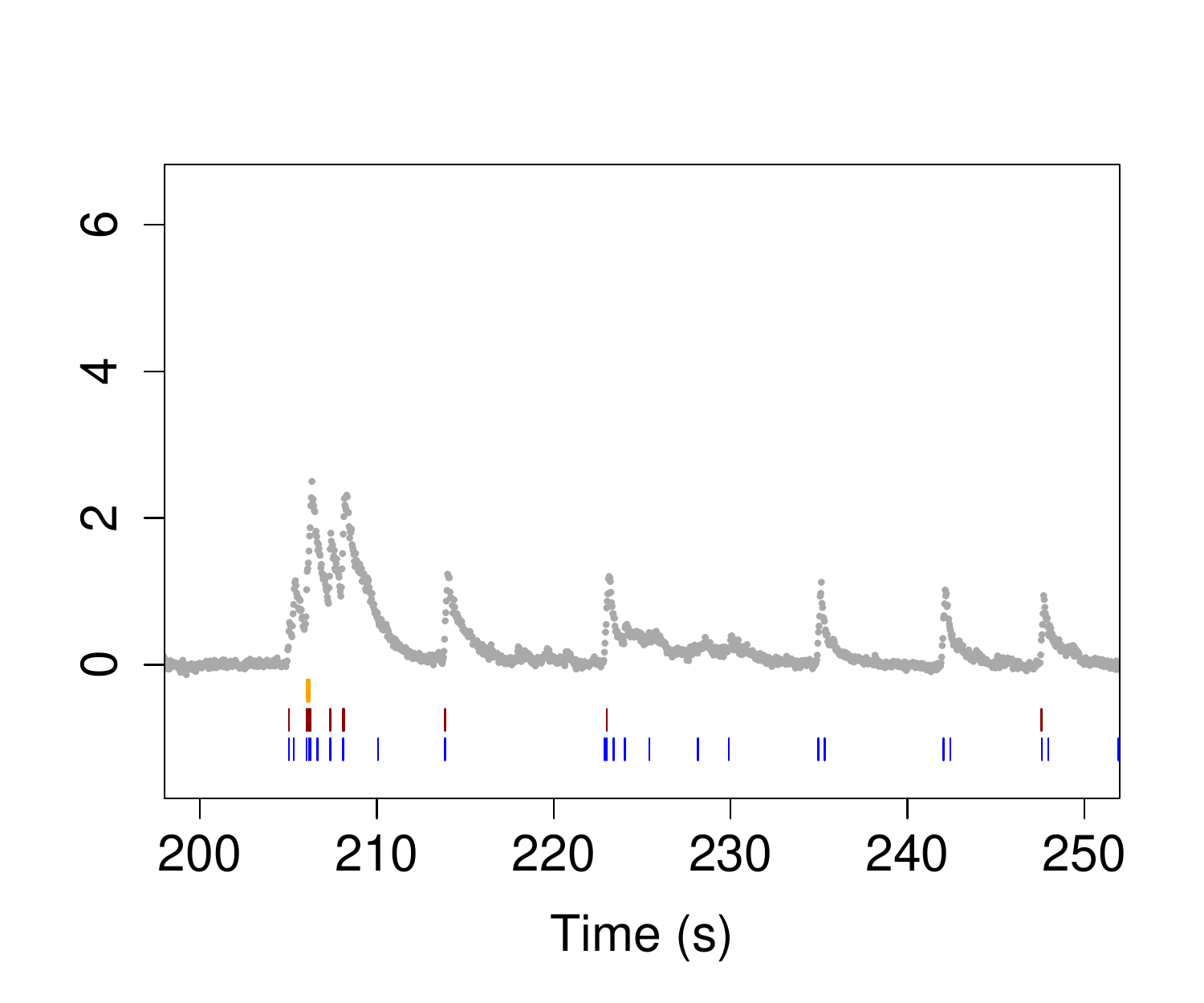}
\caption{Zoomed 128 spikes.}
\end{subfigure}
\caption{The first 10,000 timesteps from the second ROI in NWB 510221121 from the Allen Brain Observatory. Each panel displays the DF/F-transformed fluorescence (\protect\includegraphics[height=0.4em]{grey-circle}), the estimated spikes from the $\ell_{0}$ problem (\protect\includegraphics[height=0.4em]{blue}) \eqref{eq:nonconvex-nopos},  the estimated spikes from the $\ell_{1}$ problem  (\protect\includegraphics[height=0.4em]{orange}) \eqref{eq:convex-fried}, and the  estimated  spikes from post-thresholding the $\ell_1$ problem (\protect\includegraphics[height=0.4em]{darkred})  \eqref{eq:post}.  The panels display results from applying the $\ell_1$ and $\ell_0$ methods with tuning parameter $\lambda$ chosen to yield \emph{(a)}: 27 spikes for each method; \emph{(b)}: 49 spikes for each method;  and \emph{(c)}: 128 spikes for each method. The post-thresholding estimator was obtained by applying the $\ell_1$ method with $\lambda=1$, and thresholding the result to obtain 27, 49, or 128 spikes.  \emph{(d)--(f)}: As in (a)--(c), but zoomed in on 200-250 seconds.}
\label{fig:allen}
\end{figure}
\end{landscape}

For the second ROI in NWB 510221121, we applied the $\ell_1$, post-thresholded $\ell_1$, and $\ell_0$ methods to the first $10,000$ timesteps of the $DF/F$-transformed fluorescence traces. Since the data are measured at 30 Hz, this amounts to the first $333$ seconds of the recording. 
Figure~\ref{fig:allen} shows the results obtained with $\gamma = 0.981756$.
 For the $\ell_0$ and $\ell_1$ estimators,   we  chose the values of $\lambda$ in \eqref{eq:convex-fried} and \eqref{eq:nonconvex-nopos} in order to obtain 27, 49, and 128 estimated spikes. For the post-thresholded estimator \eqref{eq:post}, we set $\lambda=1$, and then selected $L$ to yield 27, 49, and 128 estimated spikes.


As in the previous subsection, we see that when faced with a large increase in fluorescence, the $\ell_1$ problem tends to estimate a very large number of spikes in quick succession. For example, when 27 spikes are estimated, the $\ell_1$ problem concentrates the estimated spikes at three points in time (Figure~\ref{fig:allen}(a)). Even when 128 spikes are estimated, the $\ell_1$ problem still seems to miss all but the largest peaks in the fluorescence data (Figure~\ref{fig:allen}(c)). Post-thresholding the $\ell_1$ estimator improves upon this issue somewhat, but spikes corresponding to smaller increases in fluorescence are still missed;  this issue can be clearly seen in Figures~\ref{fig:allen}(d)--(f), which zoom in on a smaller  time window. 

By contrast, the $\ell_0$ problem can assign an arbitrarily large increase in calcium to a single spike event. Therefore, it seems to capture most of the visible peaks in the fluorescence data when 49 spikes are estimated (Figures~\ref{fig:allen}(b) and \ref{fig:allen}(e)), and it captures all of them when 128 spikes are estimated (Figures~\ref{fig:allen}(c) and \ref{fig:allen}(f)).

  
  Though the true spike times are unknown for this data, based on visual inspection, 
  the results for the $\ell_0$ proposal seem superior to those of the $\ell_1$ and post-thresholded $\ell_1$ proposals. 
  
%

\section{Extensions}
\label{sec:extensions}

We now present two straightforward extensions to 
the optimization problem \eqref{eq:nonconvex-nopos}, for which computationally attractive algorithms along the lines of the one proposed in Section~\ref{sec:algorithm} are available.

\subsection{Estimation of the Intercept in \eqref{eq:model}}
The model for calcium dynamics considered in this paper \eqref{eq:model} allows for an intercept term, $\beta_0$. In order to arrive at  \eqref{eq:nonconvex-nopos}, we assumed that the intercept was known and (without loss of generality) equal to zero.
 However, in practice, we might want to fit the model \eqref{eq:model} without knowing the value of the intercept $\beta_0$.  In fact, in many settings, this may be of great practical importance, since the meaning of the model \eqref{eq:model} (and, for instance, the rate of exponential decay $\gamma$) is inextricably tied to the value of the intercept. 
  
We now propose a modification to the $\ell_0$ optimization problem  \eqref{eq:nonconvex-nopos} that allows for estimation of the intercept $\beta_0$. So that the resulting problem can be efficiently solved using the ideas laid out in Section~\ref{sec:algorithm}, we must ensure that given the estimated spike times, the calcium can be estimated separately between each pair of consecutive spikes.  Consequently, we must allow for a separate intercept term between each pair of consecutive spikes. This suggests the optimization problem 
\begin{equation}
\minimize{c_{1}, \ldots, c_{T}, \beta_{01}, \ldots, \beta_{0T}}{\frac12\sum_{t = 1}^{T} (y_{t} - c_{t} - \beta_{0t})^{2} + \lambda 
\sum_{t = 2}^{T} 1_{\left(c_{t} \neq \gamma c_{t-1}, \beta_{0t} \neq \beta_{0,t-1}\right)}},
\label{eq:intercept}
\end{equation}
where the indicator variable $1_{(A,B)}$ equals 1 if the event $A \cup B$ holds, and equals zero otherwise.
 Therefore, $1_{\left(c_{t} \neq \gamma c_{t-1}, \beta_{0t} \neq \beta_{0,t-1}\right)}$ equals $1$ if  the calcium concentration stops decaying or if the intercept changes.  
Note that in the solution to \eqref{eq:intercept}, the intercept is constant between adjacent timesteps, unless there is a spike.


Problem \eqref{eq:intercept} can be recast as a changepoint problem of the form \eqref{eq:cp} with 
\begin{align*}
\D(y_{a:b}) &\equiv \min \left\{ 	\frac12 \sum_{t = a}^{b} (y_{t} - c_{t} -\beta_{0t})^{2}	\right\} \mbox{ subject to } 
c_{t} = \gamma c_{t-1}, \beta_{0t} = \beta_{0,t-1}, \quad t = a+1, \ldots, b.
\end{align*}
Given $\D(y_{a:b})$, $\D(y_{a:(b+1)})$ can be updated in constant time. 
Thus, the algorithms introduced in Section~\ref{sec:algorithm} can be easily modified in order to solve \eqref{eq:intercept} for the global optimum.

\subsection{An Auto-Regressive Model with $p>1$ in  \eqref{eq:model}}

The model  \eqref{eq:model} allows for the calcium dynamics to follow a $p$th order auto-regressive process. For simplicity, this paper focused on the case of $p=1$. We now consider developing an $\ell_0$ optimization problem for the model \eqref{eq:model} with $p>1$. 

It is natural to consider the $\ell_{0}$ optimization problem 
\begin{align}
\minimize{c_{1}, \ldots, c_{T}}{\frac12\sum_{t = 1}^{T} (y_{t} - c_{t})^{2} + \lambda 
\sum_{t = p+1}^{T} 1_{\left(c_{t} \neq \sum_{i = 1}^{p}\gamma_{i} c_{t-i}\right)}}.
\label{eq:l0p-doesntwork}
\end{align}
However,   \eqref{eq:l0p-doesntwork} cannot be expressed in the form \eqref{eq:cp}:  the penalty in \eqref{eq:l0p-doesntwork} induces a dependency in the calcium that spans more than two timesteps, 
so that the calcium at a given timestep may depend on the calcium prior to the most recent spike. 
As a result, \eqref{eq:l0p-doesntwork} is computationally intractable.


Instead, we consider a changepoint detection problem of the form \eqref{eq:cp} with cost function
\begin{align*}
\D(y_{a:b}) &\equiv \min \left\{ 	\frac12 \sum_{t = a}^{b} (y_{t} - c_{t})^{2}	\right\} \mbox{ subject to } 
c_{t} = \sum_{i = 1}^{p}\gamma_{i} c_{t-i}, \quad t = a + p, \ldots, b.
\end{align*}
Thus, the calcium follows a $p$th order auto-regressive model between any pair of spikes; furthermore, once a spike occurs, the calcium concentrations are reset completely. That is, the calcium after a spike is not a function of the calcium before a spike. Consequently,  it is straightforward to develop a fast algorithm for solving this changepoint detection problem for the global optimum,  using ideas detailed in Section~\ref{sec:algorithm}. 

In particular, a popular model for calcium dynamics assumes that between any pair of spikes, the calcium can be well-approximated by the difference between two exponentially-decaying functions \citep{brunel2003determines,mazzoni2008encoding,cavallari2016comparison,volgushev2015identifying}. This would perhaps be a better model for the data from the Allen Brain Observatory, in which  increases in fluorescence due to a  spike occur over the course of a few timesteps, rather than instantaneously (see Figure~\ref{fig:allen}). 
This ``difference of exponentials" models falls directly within the framework of \eqref{eq:model} with $p=2$, and hence could be handled handled using the changepoint detection problem just described.


%
%

\section{Discussion}
\label{sec:discussion}

In this paper, we considered solving the seemingly intractable $\ell_0$ optimization problem \eqref{eq:nonconvex-nopos} corresponding to the model \eqref{eq:model}. By recasting \eqref{eq:nonconvex-nopos}  as a changepoint detection problem, we were able to derive an algorithm to solve  \eqref{eq:nonconvex-nopos} for the global optimum in expected linear time.
   It should be possible to develop an even more efficient algorithm for solving \eqref{eq:nonconvex-nopos} that exploits recent algorithmic developments for changepoint detection \citep{johnson2013dynamic,maidstone2016optimal, hocking2017log}; we leave this as an avenue for future work.

We have shown in this paper that solving the $\ell_0$ optimization problem \eqref{eq:nonconvex-nopos} leads to more accurate spike event detection than solving the $\ell_1$ optimization problem \eqref{eq:convex-fried} proposed by \cite{friedrichfast2017}. Indeed, this finding is intuitive: the $\ell_1$ penalty and positivity constraint in  \eqref{eq:convex-fried} serves as a exponential prior on the increase in calcium at any given time point, and thereby effectively limits the amount that calcium can increase in response to a spike event.   By contrast, the $\ell_0$ penalty in \eqref{eq:nonconvex-nopos} is completely agnostic to the amount by which a spike event increases the level of calcium. Consequently, it can allow for an arbitrarily large (or small) increase in fluorescence as a result of a spike event. 

While approximations to the solution to the $\ell_{0}$ problem \eqref{eq:nonconvex-nopos} are possible \citep{de2011deconvolution, de2014sparse, hugelier2016sparse, scott1974cluster, olshen2004circular, fryzlewicz2014wild, friedrichfast2017}, there is no guarantee that such approaches will yield an attractive local optimum on a given data set. In this paper, we completely bypass this concern by solving the $\ell_{0}$ problem for the global optimum.

In this paper, we have focused on the empirical benefits of the $\ell_0$ problem \eqref{eq:nonconvex-nopos} over the $\ell_1$ problem \eqref{eq:convex-fried}. However, it is natural to wonder whether these empirical benefits are backed by statistical theory. Conveniently, both the $\ell_0$ and $\ell_1$ optimization problems are very closely-related to problems that have been well-studied in the statistical literature from a theoretical standpoint. In particular, in the special case of $\gamma=1$, the $\ell_0$ problem \eqref{eq:nonconvex-nopos} was extensively studied in \citet{yao1989least} and \cite{boysen2009consistencies}. 
Furthermore, when $\gamma=1$,  the $\ell_{1}$ problem \eqref{eq:nonconvex-nopos} is very closely-related to the \emph{fused lasso} optimization problem, 
$$
\minimize{c_1,\ldots,c_T}{  \frac{1}{2} \sum_{t=1}^T \left( y_t - c_t \right)^2 + \lambda \sum_{t=2}^T | c_t -  c_{t-1} |},$$
which has also been extremely well-studied  \citep{tsrz2005, mammen1997locally, davies2001local, rinaldo2009properties, harchaoui2010multiple, qian2012pattern, rojas2014change, lin2016approximate, dalalyan2017prediction}. 
 However, we leave a formal theoretical analysis of the relative merits of   \eqref{eq:nonconvex-nopos} and \eqref{eq:convex-fried}, in terms of $\ell_2$ error bounds and spike recovery properties,   to future work.

Our \verb=R=-language software for our proposal is available on \verb=CRAN= in the package\\ \verb=LZeroSpikeInference=. Instructions for running this software in \verb=python= can be found at \url{https://github.com/jewellsean/LZeroSpikeInference}.

\section*{Acknowledgments}

We are grateful to  three reviewers and an associate editor for helpful comments that improved the quality of this work. 
We thank Michael Buice and Kyle Lepage at the Allen Institute for Brain Science for conversations that motivated this work.  Johannes Friedrich at Columbia University provided  assistance in using his software to solve the $\ell_1$ problem \eqref{eq:convex-fried}. Sean Jewell received funding from the Natural Sciences and Engineering Research Council of Canada. This work was partially supported by NIH Grant DP5OD009145 and NSF CAREER Award DMS-1252624 to Daniela Witten.

\clearpage
\appendix
\section{Proof of Propositions}
\label{sec:appendix}

\subsection{Proof of Proposition~\ref{prop:equiv}}

\begin{proof}

The first sentence follows by inspection. To establish the second sentence, we observe that the cost
\begin{align*}
 \D(y_{a:b}) \equiv \underset{ c_{a}, c_t = \gamma c_{t-1}, \; t=a+1,\ldots,b}{\min}
  \left\{  \frac{1}{2} \sum_{t=a}^{b} \left( y_t - c_t \right)^2 \right\}
\end{align*}
can be rewritten by direct substitution of the constraint as
\begin{align*}
 \D(y_{a:b}) =  \underset{ c_a }{\min}
  \left\{  \frac{1}{2} \sum_{t=a}^{b} \left( y_t - \gamma^{t-a}c_a \right)^2 \right\}.
\end{align*}
This is a least squares problem and is minimized at
\begin{align*}
\hat{c}_{a} = \frac{	\sum_{t=a}^{b} y_{t}\gamma^{t-a}}{ \sum_{t=a}^{b} \gamma^{2(t-a)}},
\end{align*}
which implies that
\begin{align*}
\D(y_{a:b}) = \frac{1}{2} \sum_{t=a}^{b} \left( y_t - \gamma^{t-a}\hat{c}_a \right)^2,
\end{align*}
and furthermore that for $a < t \leq b$ the fitted values are $\hat{c}_{t} = \gamma \hat{c}_{t-1}$. Applying this argument to each segment gives the result stated in Proposition~\ref{prop:equiv}.

\end{proof}

\subsection{Proof of Proposition~\ref{prop:segment-soln}}

\begin{proof}
The first equation follows by expanding the square for the final form of $\D(y_{a:b})$ in the proof of Proposition~\ref{prop:equiv}. Given $\D(y_{a:b})$ we can calculate $\D(y_{a:(b+1)})$ in constant time by storing $\sum_{t=a}^{b} \frac{y_{t}^{2}}{2}$ and $\sum_{t=a}^{b} y_{t}\gamma^{t-a}$, and updating each of these sums for the new data point $y_{b+1}$; we use a closed form expression to calculate $\sum_{t=a}^{b+1} \gamma^{2(t-a)}$. With each of these quantities stored, $\D(\cdot)$ and $\C(\cdot)$ are updated in constant time.

\end{proof}

\section{Choosing $\lambda$ and $\gamma$}
\label{subsec:cv}

Recall that in  \eqref{eq:nonconvex-nopos}, the parameters $\lambda$ and $\gamma$ are unknown. 
The nonnegative parameter $\lambda$  controls the trade-off between the number of estimated spike events and the quality of the estimated calcium fit to the observed fluorescence. The parameter $\gamma, 0<\gamma < 1$, controls the rate of exponential decay of the calcium. We consider two approaches for choosing $\gamma$ and $\lambda$. 

\subsection{Approach 1} 

To estimate $\gamma$, we manually select a segment $y_{a:b}$ that, based on visual inspection, appears to  exhibit exponential decay. We then estimate $\gamma$ as
\begin{align*}
\hat{\gamma} =	\argmin{\gamma}{\D\left(y_{a:b}\right) } = \argmin{\gamma}{ \underset{c_{a}, c_{t} = \gamma c_{t-1}, t = a + 1, \ldots, b}{\min} \left\{ 	\frac12\sum_{t=a}^{b} (y_{t} - c_{t})^{2}	\right\}}.
\end{align*}
This can be done via numerical optimization.

Next, given $\gamma$, we select $\lambda$ via cross-validation. For each value of $\lambda$ that we consider, we solve \eqref{eq:cp} on a training set, and then evaluate the mean squared error (MSE) on a hold-out set. Details are provided in Algorithm~\ref{alg:cv}.

\subsection{Approach 2}

 \cite{pnevmatikakis2013bayesian}, \cite{friedrich2016fast}, and \cite{friedrichfast2017} propose to select the exponential decay parameter $\gamma$ based on the autocovariance function, and to choose the tuning parameter $\lambda$ such that $||y-\hat{c}||_{2} \leq \sigma \sqrt{T}$, where the standard deviation $\sigma$ is estimated through the power spectral density of $y$, and $T$ is the number of timepoints. We refer the reader to \cite{friedrichfast2017} and \cite{pnevmatikakis2016simultaneous} for additional details.

\begin{algorithm}[h!]
    \SetKwInOut{Initialize}{Initialize} 
    \SetKwInOut{Output}{Output}
	\Initialize{ Candidate tuning parameter values $\lambda_{1}, \ldots, \lambda_{M}$;  a fixed value $\gamma$ for the rate of exponential decay; a matrix $\text{cvMSE} \in\R^{M \times 2}$ to store the cross-validated MSEs.}
	
	\ForEach{\text{fold} in $1,2$}{
	
	Assign odd timesteps to the training set and even timesteps to the test set for the first fold, and vice-versa for the second fold. Note that $\mathrm{card}(y^{\text{train}})=\mathrm{card}(y^{\text{test}})=T/2$.

	\ForEach{$m=1,\ldots,M$ }{

	Solve \eqref{eq:cp} on the  training set $y^{\text{train}}$ with tuning parameter values $\lambda_m$ and $\gamma^{2}$, in order to obtain an estimate of the changepoints $\tau_1,\ldots,\tau_k$. Set $\tau_0=0$ and $\tau_{k+1}=T/2$. \\
	Average adjacent fitted values in order to obtain predictions on the test set $y^{\text{test}}$,
	\begin{align*}
		\hat{c}^{\text{test}} &= \frac{\hat{c}^{\text{train}}_{1:(T/2-1)} + \hat{c}^{\text{train}}_{2:(T/2)}}{2}.
	\end{align*}\\
	Calculate and store the test set MSE,
	\begin{align*}
		\text{cvMSE}_{m, \text{fold}} &= \frac{2}{T} \sum_{t = 1}^{T/2} \left( y^{\text{test}}_{t} - \hat{c}_{t}^{\text{test}}\right)^{2}.
	\end{align*}
	\\ 
	}}
	
	Average the test set MSE over folds,
	\begin{align*}
	\overline{\text{cvMSE}}_{m} =  \frac12 \left( \text{cvMSE}_{m, 1} + \text{cvMSE}_{m, 2} \right), \quad
	\end{align*}\\
	Calculate $\hat{m} = \argmin{m}{\overline{\text{cvMSE}}_{m}}$.\\
	Calculate the standard error of the test set MSE over folds,
	\begin{align*}
	\text{se}\left({\text{cvMSE}}\right)_{m} &= \sqrt{\frac{\left( \text{cvMSE}_{m, 1} - 	\overline{\text{cvMSE}}_{m}	\right)^{2} + \left( \text{cvMSE}_{m, 2} - 	\overline{\text{cvMSE}}_{m}	\right)^{2}}{2}}, \mbox{ for } m=1,\ldots,M.
	\end{align*}\\
	Calculate $$m^* = \max\left\{m: \overline{\text{cvMSE}}_{m} \leq \overline{\text{cvMSE}}_{\hat{m}} + \text{se}\left({\text{cvMSE}}\right)_{\hat{m}}\right\}.$$

	\Output{The value
	 $\lambda_{\hat{m}}$ that 
	minimize the cross-validated MSE, and the values
	$\lambda_{m^*}$ selected 
	based on the one-standard-error rule \citep{elemstatlearn}.}
    \caption{A Cross-Validation Scheme for Choosing $\lambda$ \eqref{eq:nonconvex-nopos}}\label{alg:cv}
\end{algorithm}

\clearpage
\section{A Greedy Approach for Approximating the Solution to a Non-Convex Problem}
\label{sec:heur}


\cite{friedrichfast2017} consider a variant of the optimization problem \eqref{eq:convex-fried},
\begin{equation}
\minimize{c_1,\ldots,c_T, s_2,\ldots,s_T}{  \frac{1}{2} \sum_{t=1}^T \left( y_t - c_t \right)^2} \mbox{ subject to } s_t = c_t - \gamma c_{t-1} \geq s_{min} \mbox{ or } s_{t} = 0,
   \label{eq:nonconvex-smin}
   \end{equation}
obtained from \eqref{eq:convex-fried} by setting $\lambda = 0$, and changing the convex positivity constraint to the non-convex constraint 
that $s_t$ lies within a non-convex set. Like \eqref{eq:nonconvex-nopos}, \eqref{eq:nonconvex-smin} is non-convex. 
\cite{friedrichfast2017}  do not attempt to solve \eqref{eq:nonconvex-smin} for the global optimum; instead, they provide a heuristic modification to their algorithm for solving \eqref{eq:nonconvex-nopos}, which is intended to approximate the solution to \eqref{eq:nonconvex-smin}. 

\begin{landscape}

\begin{figure}
\centering
\begin{subfigure}[b]{0.4\textheight}
\centering
\includegraphics[scale=.5]{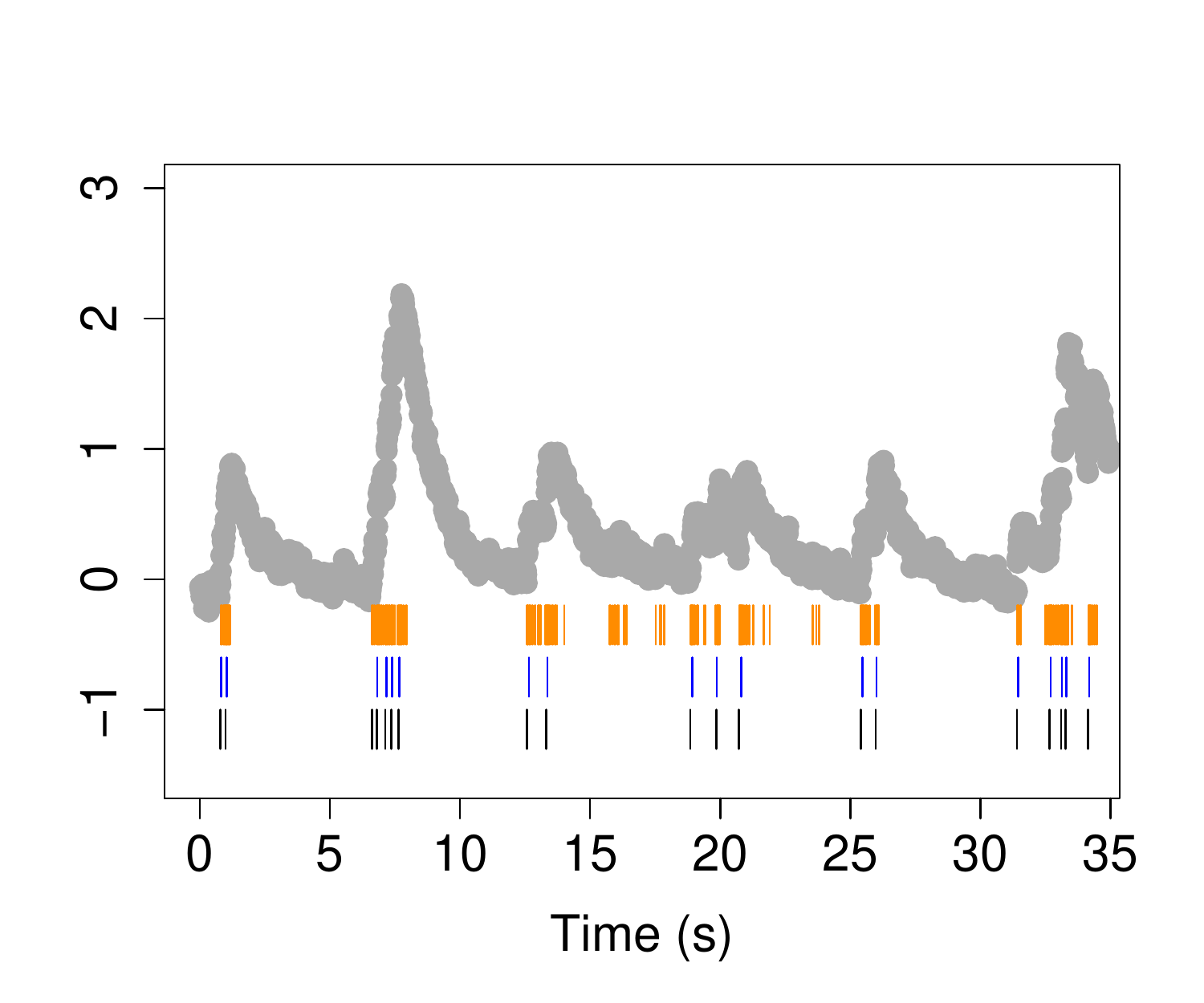} 
\caption{$s_{min} = 10^{-8}$}
\end{subfigure}
\begin{subfigure}[b]{0.4\textheight}
\centering
\includegraphics[scale=.5]{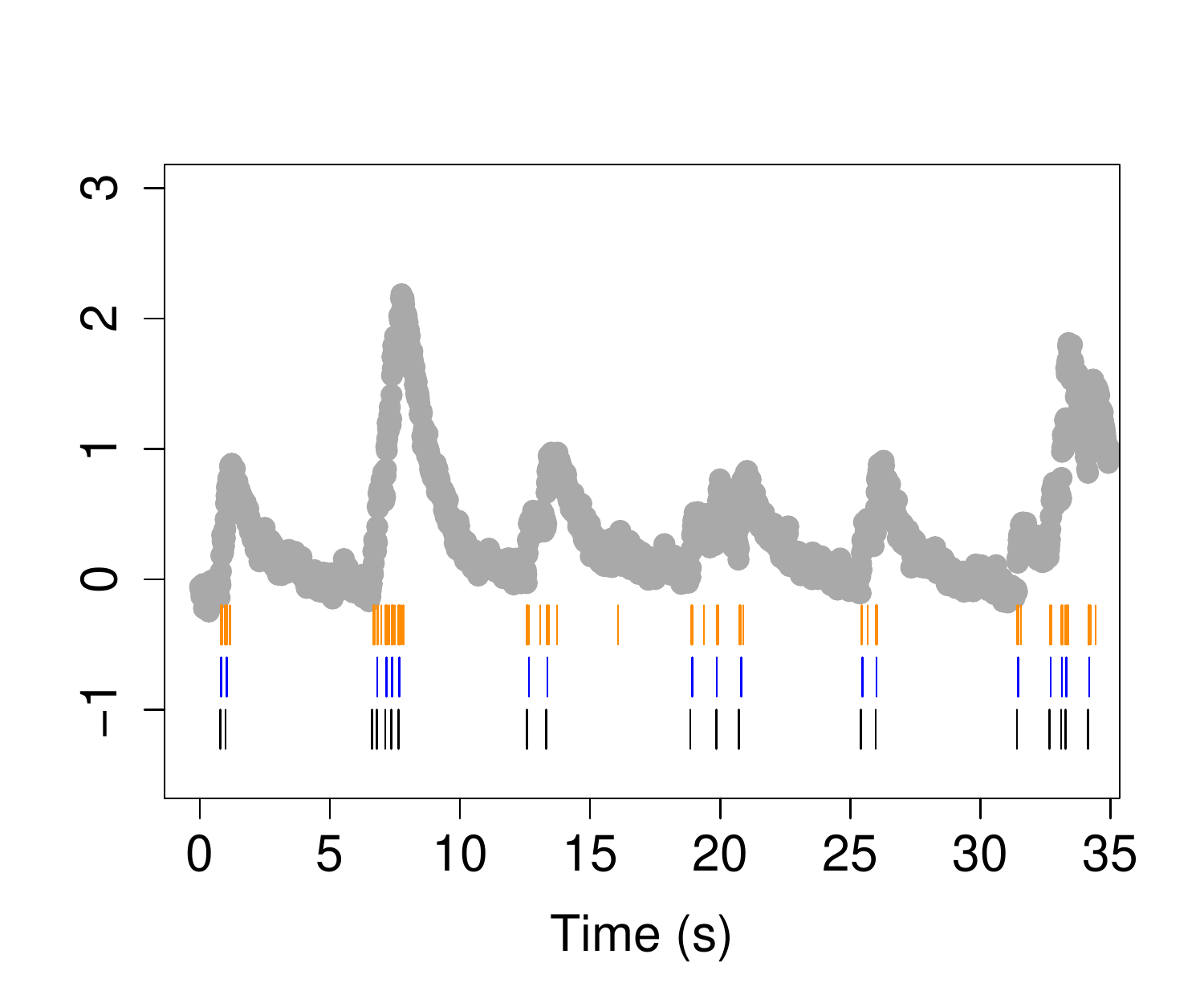} 
\caption{$s_{min} = 0.1$}
\end{subfigure}
\begin{subfigure}[b]{0.4\textheight}
\centering
\includegraphics[scale=.5]{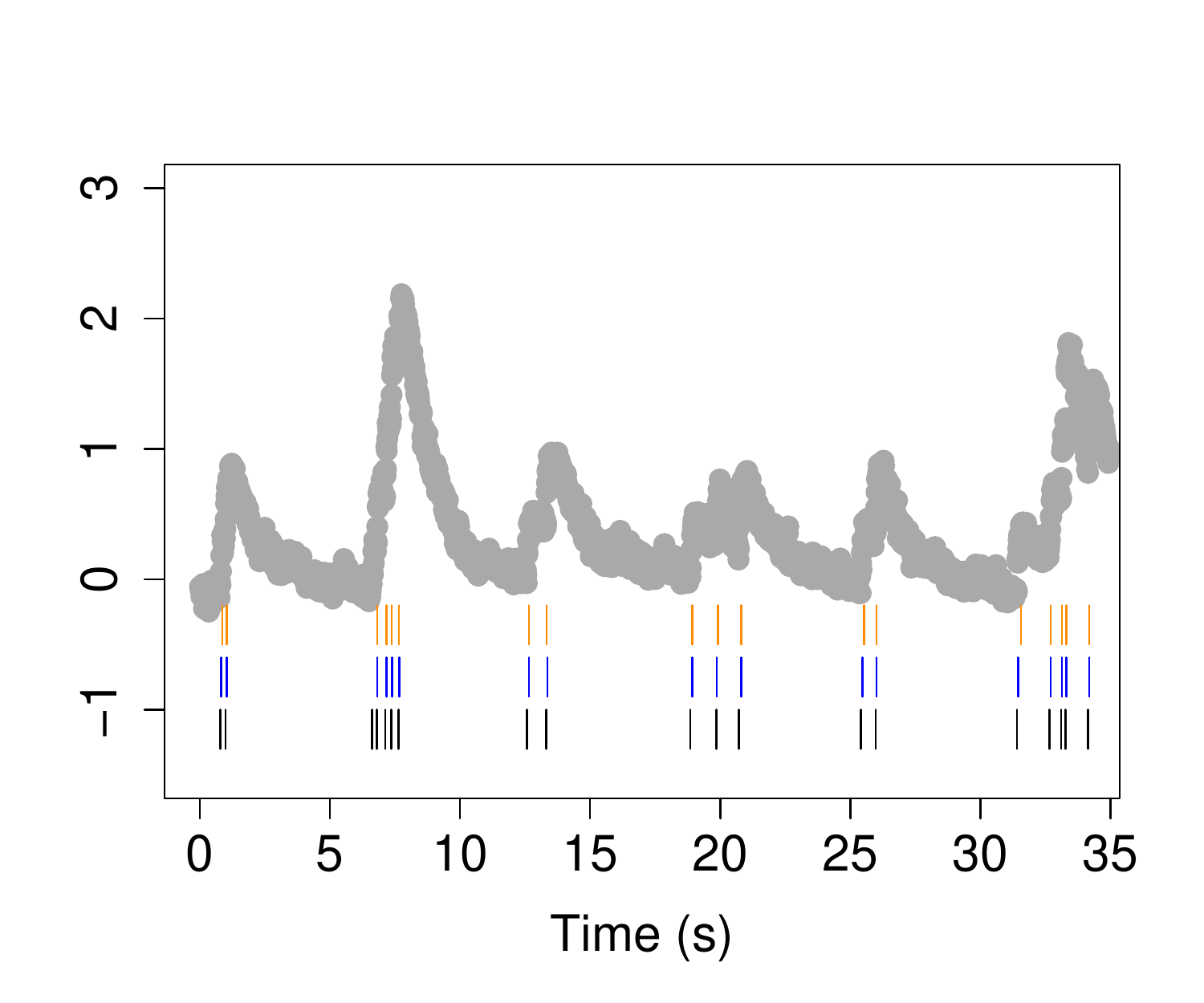}  
\caption{$s_{min} = 0.3$}
\end{subfigure}%

\begin{subfigure}[b]{0.4\textheight}
\centering
\includegraphics[scale=.5]{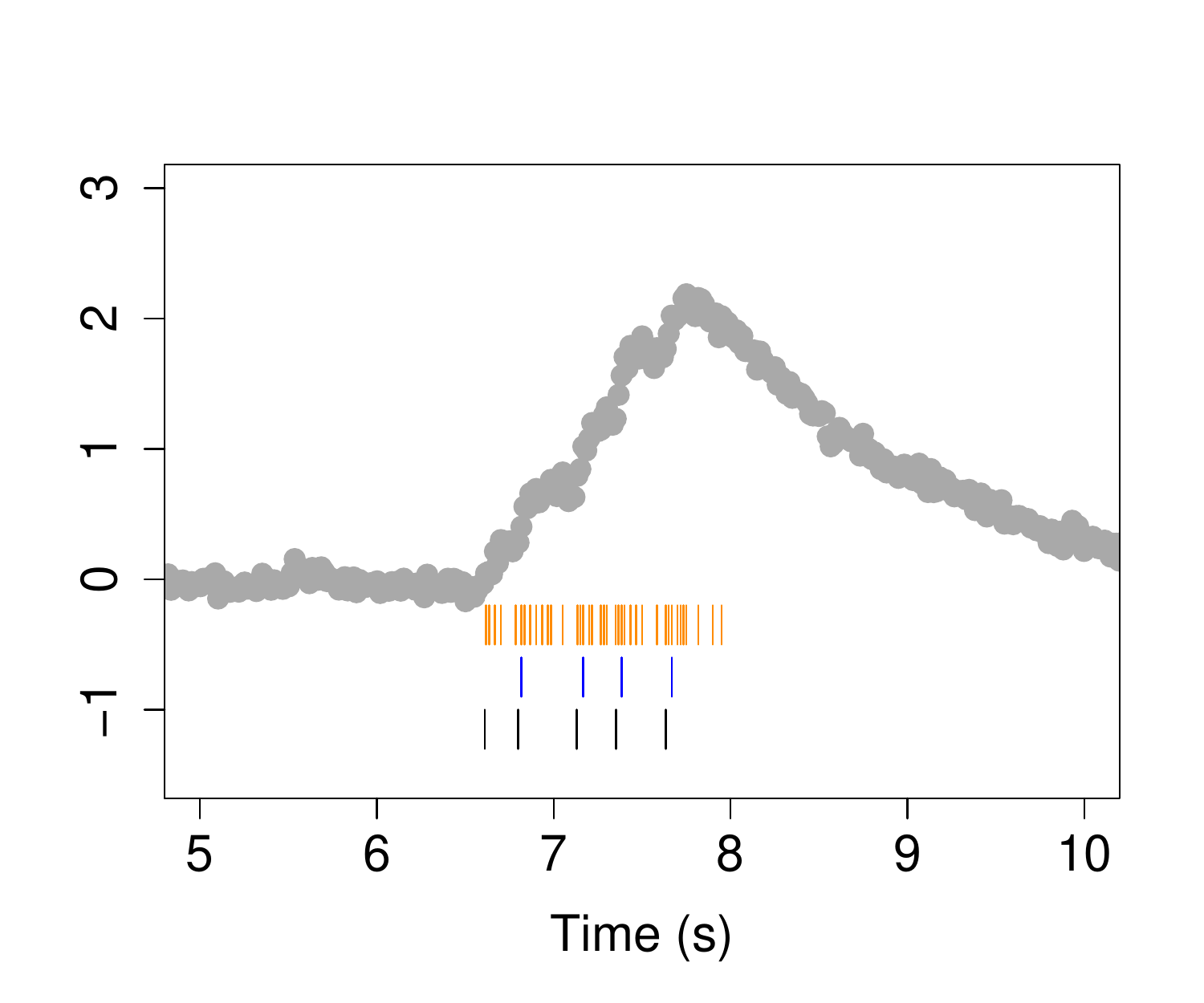} 
\caption{Zoomed $s_{min} = 10^{-8}$}
\end{subfigure}
\begin{subfigure}[b]{0.4\textheight}
\centering
\includegraphics[scale=.5]{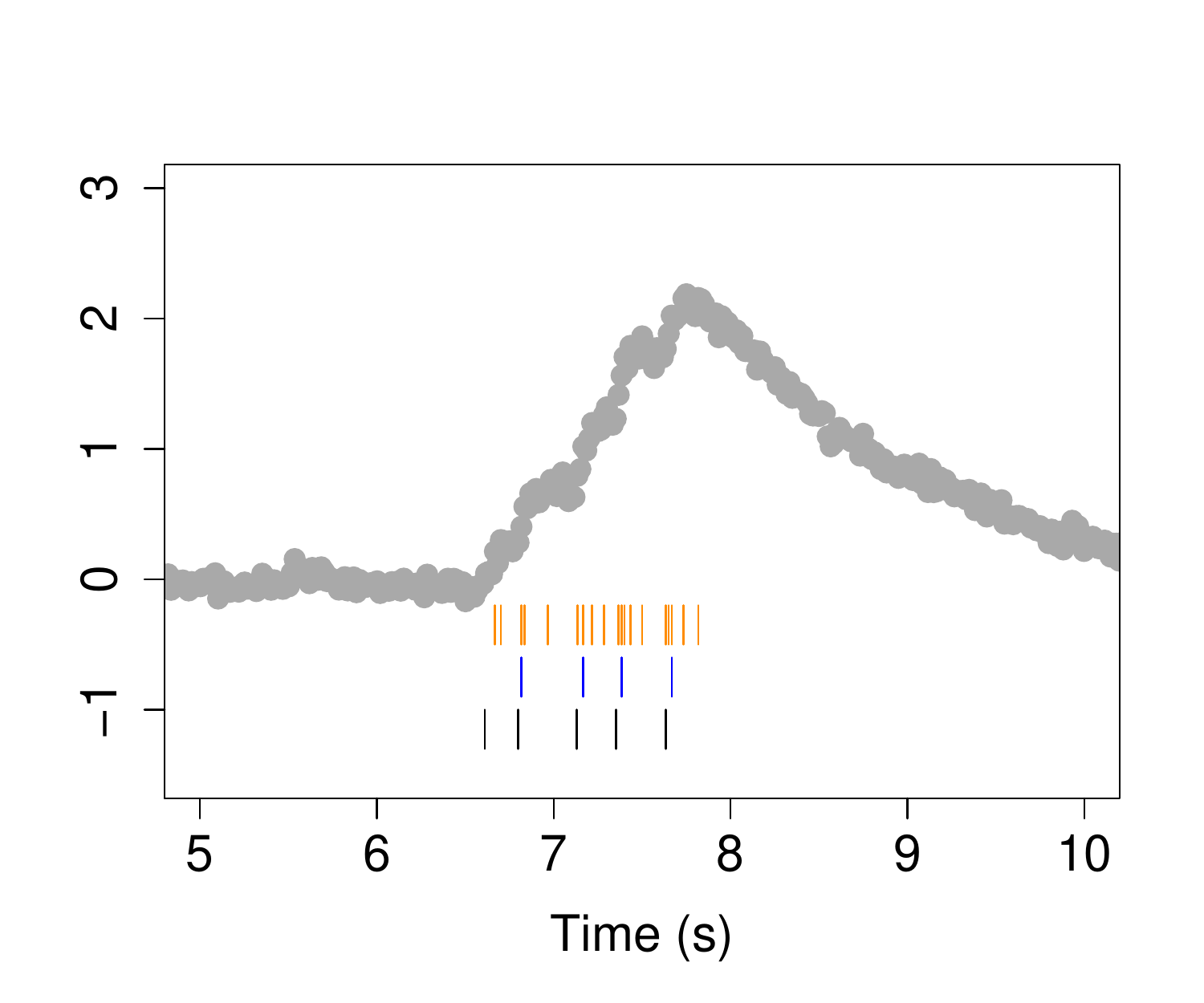} 
\caption{Zoomed $s_{min} = 0.1$}
\end{subfigure}
\begin{subfigure}[b]{0.4\textheight}
\centering
\includegraphics[scale=.5]{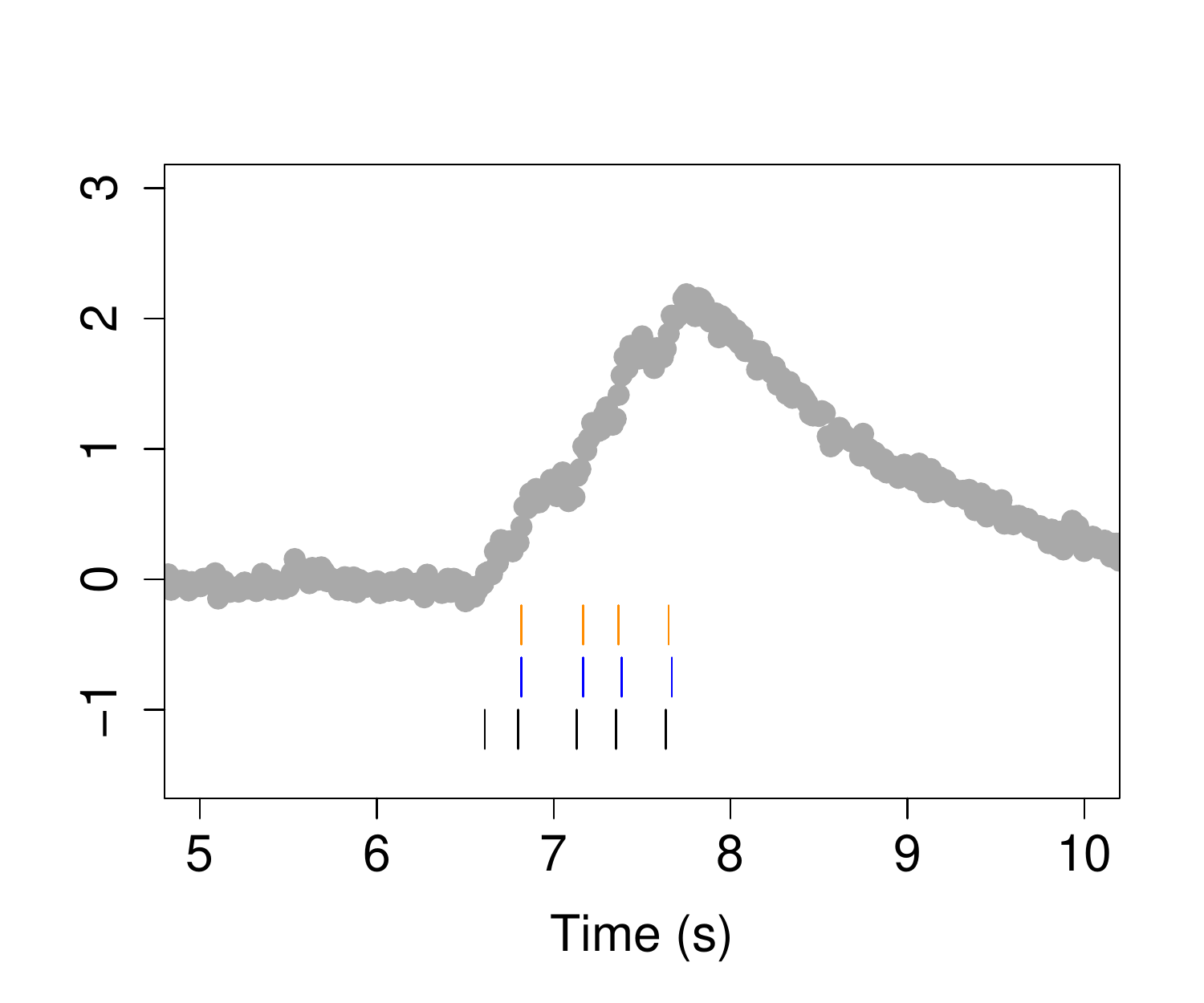} 
\caption{Zoomed $s_{min} = 0.3$}
\end{subfigure}
\caption{Spike detection for  cell 2002 of the \citet{chen2013ultrasensitive} data. In each panel, the observed fluorescence (\protect\includegraphics[height=0.4em]{grey-circle}) and true spikes (\protect\includegraphics[height=0.3em]{black}) are displayed. Estimated  spikes from problem \eqref{eq:nonconvex-smin} are shown in (\protect\includegraphics[height=0.3em]{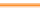}), and the estimated spikes from the $\ell_{0}$ problem \eqref{eq:nonconvex-nopos} with $\lambda = 0.6$ are shown in (\protect\includegraphics[height=0.4em]{blue}). Times $0s-35s$ are shown in the top row; the second row zooms in on times $5s-10s$ to illustrate behavior around a large increase in calcium concentration. Columns correspond to different values of $s_{min}$. 
}
\label{fig:smin}
\end{figure}
\end{landscape}

Figure~\ref{fig:smin} illustrates the behavior of this approximate algorithm when  applied to the same data as in Figure~\ref{fig:chen1}. We set    $\gamma = 0.9864405$, and considered three values of $s_{min}$. When $s_{min}=10^{-8}$ and $s_{min}=0.1$, in panels (a) to (b), too many spikes are estimated.  But when $s_{min}=0.3$, in panel (c), the solution to \eqref{eq:nonconvex-smin} is very similar to the solution to \eqref{eq:nonconvex-nopos} with $\lambda = 0.6$. Both almost perfectly recover  the ground truth spikes. Therefore, in this example, the approximate algorithm of \cite{friedrichfast2017} for solving \eqref{eq:nonconvex-smin} performs quite well.

However, \eqref{eq:nonconvex-smin} is a non-convex problem, and the approximate algorithm of \cite{friedrichfast2017}  is not guaranteed to find the global minimum. 
In fact, we can see that on the data shown in Figure~\ref{fig:smin}, this approximate algorithm does not find the global optimum. When applied with $s_{min} = 0.3$, the approximate algorithm yields an objective value of $8.57$. By contrast, our algorithm for solving \eqref{eq:nonconvex-nopos} yields a solution that is feasible for \eqref{eq:nonconvex-smin}, and which results in a value of $7.86$ for the objective of \eqref{eq:nonconvex-smin}. We emphasize that this is quite remarkable: even though the algorithm proposed in Section~\ref{sec:algorithm} solves \eqref{eq:nonconvex-nopos} and not \eqref{eq:nonconvex-smin}, \emph{it nonetheless yields a solution that is closer to the global optimum of \eqref{eq:nonconvex-smin} than does the approximate algorithm of \cite{friedrichfast2017}, which is intended to solve \eqref{eq:nonconvex-smin}}. 

In many cases, the greedy algorithm of \cite{friedrichfast2017} for solving \eqref{eq:nonconvex-smin} might yield good results that are near the global optimum of \eqref{eq:nonconvex-smin}, and potentially even near the global optimum of \eqref{eq:nonconvex-nopos}. However, there is no guarantee that this algorithm will yield a ``good" local optimum on any given data set. 
By contrast, in this paper we have proposed an elegant and efficient algorithm for exactly solving the $\ell_{0}$ problem \eqref{eq:nonconvex-nopos}.



\clearpage

\bibliographystyle{imsart-nameyear}
\bibliography{paper-ref,mega}

\end{document}